\documentclass[12pt]{article}
\usepackage[utf8]{inputenc} 
\RequirePackage{amsthm,amsmath,amssymb}
\usepackage{graphicx,psfrag,epsf}
\usepackage{enumerate}
\usepackage{natbib}
\usepackage{url} 
\usepackage{mathtools}
\usepackage{float}
\usepackage[caption = false]{subfig}
\usepackage{xcolor}
\usepackage[colorlinks = true,
            linkcolor = blue,
            urlcolor  = blue,
            citecolor = blue]{hyperref}
 \usepackage{MnSymbol,wasysym}
\usepackage{bbold}

\newcommand{\blind}{1}

\usepackage{mathbbol}

\usepackage{stackrel}

\DeclareMathOperator*{\argmin}{arg\,min}
\DeclareMathOperator*{\argmax}{arg\,max}

\newcommand{\dens}{\texttt{p}}

\def\by{y}
\def\bx{X}

\def\like{{\cal L}}
\def\llike{{\ell}}

\def\iid{\mathop{\sim}_{i.i.d.}}

\def\ximis{\bx_{i,{\rm mis}}}
\def\xiobs{\bx_{i,{\rm obs}}}
\def\xmis{\bx_{{\rm mis}}}
\def\xobs{\bx_{{\rm obs}}}
\def\xa{\bx^{(\rm mis)}}
\def\xb{\bx^{({\rm obs})}}

\def\xm{x_{{\rm mis}}}
\def\xo{x_{{\rm obs}}}

\def\nmdist{\mathcal{N}}
\def\bslope{\hat{\beta}_{\rm SLOPE}}

\def\sig2{\sigma^2}

\def\iden{\mathbb{1}}

\usepackage{algorithmic}
\usepackage[Algorithm,ruled]{algorithm}

\usepackage{capt-of}

\DeclareOldFontCommand{\rm}{\normalfont\rmfamily}{\mathrm}
\DeclareOldFontCommand{\sf}{\normalfont\sffamily}{\mathsf}
\DeclareOldFontCommand{\tt}{\normalfont\ttfamily}{\mathtt}
\DeclareOldFontCommand{\bf}{\normalfont\bfseries}{\mathbf}
\DeclareOldFontCommand{\it}{\normalfont\itshape}{\mathit}
\DeclareOldFontCommand{\sl}{\normalfont\slshape}{\@nomath\sl}
\DeclareOldFontCommand{\sc}{\normalfont\scshape}{\@nomath\sc}

\newtheorem{remark}{Remark}
\newtheorem{prop}{Proposition}
\usepackage[colorinlistoftodos,textwidth=2.3cm]{todonotes}

\usepackage{makecell,bbm}
\usepackage{natbib}
\usepackage{multirow}
\newcommand\MyBox[2]{
  \fbox{\lower0.75cm
    \vbox to 1.7cm{\vfil
      \hbox to 1.7cm{\hfil\parbox{1.4cm}{#1\\#2}\hfil}
      \vfil}%
  }%
}

\usepackage{titling}
\thanksmarkseries{arabic}

\begin{document}
\def\spacingset#1{\renewcommand{\baselinestretch}%
{#1}\small\normalsize} \spacingset{1}

\date{October 2019}


\if1\blind
{
  \title{\bf Adaptive Bayesian SLOPE -- High-dimensional Model Selection with Missing Values}
        \author{Wei Jiang\thanks{Inria XPOP and CMAP, École Polytechnique, France} \and Ma{\l}gorzata Bogdan\thanks{University of Wroclaw, Poland and Lund University, Sweden} \and Julie Josse\footnotemark[1] \and B{\l}a{\.z}ej Miasojedow\thanks{University of Warsaw, Poland} \and Veronika Ro{\v{c}}kov{\'a}\thanks{University of Chicago Booth School of Business, USA} \and TraumaBase$^\circledR$  Group\thanks{Hôpital Beaujon, APHP, France}}
  \maketitle
} \fi

\if0\blind
{
  \bigskip
  \bigskip
  \bigskip
  \begin{center}
    {\LARGE\bf Adaptive Bayesian SLOPE -- High-dimensional Model Selection with Missing Values}
\end{center}
  \medskip
} \fi

\begin{abstract}
{
We consider the problem of variable selection in high-dimensional settings  with missing observations among the covariates.
To address this relatively understudied problem, we propose a new synergistic procedure -- adaptive Bayesian SLOPE -- which effectively combines  the SLOPE method (sorted $l_1$  regularization) 
together with the Spike-and-Slab LASSO method. We position our approach within a Bayesian framework which allows for simultaneous variable selection and  parameter estimation, despite the missing values. 
As with the Spike-and-Slab LASSO,  the coefficients are regarded as arising from a hierarchical model consisting of two groups: (1) the spike for the inactive and  (2) the slab for the active. However, instead of assigning independent spike priors for each covariate, here we deploy a joint ``SLOPE'' spike prior which takes into account  the ordering of  coefficient magnitudes in order to control for false discoveries.
 Through extensive simulations, we demonstrate satisfactory performance in terms of power, FDR and estimation bias  under a wide range of  scenarios. Finally, we analyze a real dataset consisting of   patients from Paris hospitals who underwent  severe trauma, where we show excellent performance in predicting platelet levels. 
 Our methodology has been implemented in C++  and wrapped into an R package {\sc ABSLOPE} for public use.}

\end{abstract}

\noindent%
{\it Keywords:}  incomplete data, FDR control, penalized regression, spike and slab prior, stochastic approximation EM, health data

\vfill

\newpage

\spacingset{0.99} 
\tableofcontents
\spacingset{1.45}  
\section{Introduction}\label{sec:intro}
{
The selection of variables from high-dimensional data is an ubiquitous problem in many contemporary data applications. In molecular genetics, for example,  a vast number of predictors is available but only a few are deemed relevant for explaining biological phenomena. 
The LASSO \citep{lasso}, now a default penalized likelihood method, has  proved itself to be successful at simultaneously  estimating parameters and   covariate sets.
While LASSO possesses nice theoretical guarantees,  it may lead to  false discoveries \citep{su2017false} and it allows to identify the true model only under rather  strict ``irrepresentability'' conditions \citep{wainwright2009, tardivel2018sign}. 
The  adaptive LASSO variant \citep{zou2006adaptive}, which instead uses   a weighted  $\ell_1$ penalty (adjusting regularization  based on some initial estimates of regression coefficients),   reduces   bias in estimation and can be consistent for variable selection even when the irrepresentability condition is not satisfied 
(see \textit{e.g.} \citet{Fan2014, tardivel2018sign, rejchel2019}). However, performance properties of adaptive LASSO still  rely heavily on the weight function and tuning parameters, whose  optimal choices  depend on  unknown aspects of the estimation problem such as signal magnitude or   sparsity. 

More recently, \citet{spikeslablasso} developed the Spike-and-Slab LASSO (SSL) procedure which bridges the default  penalized likelihood approach (the LASSO) and the default Bayesian  variable selection approach (spike-and-slab). 
In SSL, the penalty function arises from a fully Bayes spike-and-slab formulation and, as such, exerts self-adaptation properties with less hyper-parameter tuning required. In addition, SSL alleviates over-shrinkage of important signals by providing enough prior support for large effects.
Theoretical results and simulations reported in \citet{spikeslablasso} and \citet{ssl2} show that SSL attains near rate-minimax convergence (for the posterior mode {\em as well as} the entire posterior)  and performs very well even when the columns in the design matrix are strongly correlated. 

In this article we build on the Spike-and-Slab LASSO framework by incorporating aspects of the Sorted L-One Penalized Estimator (SLOPE) method of \citet{bogdan2015}. The main motivation behind SLOPE was the control of the False Discovery Rate (FDR).  Controlling FDR  is  one of the central goals of many methodological developments in multiple regression (see \textit{e.g.} \citet{barber2015controlling, panning}). Compared to methods aiming at perfect signal recovery,  controlling for FDR is more liberal as it allows for some small number of  mistakes. As a result, this leads to substantial gains in power and in prediction improvements   when the signal is weak.  As shown in \citet{bogdan2015}, SLOPE controls for FDR when the design matrix is orthogonal. Moreover,   \citet{su2016} and \citet{Bellec1} showed that, contrary to the LASSO,  SLOPE allows one to achieve the exact minimax convergence rate for regression coefficients in sparse high dimensional  regression. However, similarly as  with the LASSO,  it is challenging to attain  good prediction and, at the same time, good  variable selection  with SLOPE in finite samples.
Large amounts of shrinkage, needed to  keep FDR small, result in large estimation bias of important regression coefficients and  thereby  poor estimation. One practical remedy, suggested by \citet{bogdan2015, grpSLOPE},  is proceeding in two steps: \textit{i)} using SLOPE to detect relevant predictors; \textit{ii)} applying  standard least-squares  with selected predictors for estimation.
This two-step approach allows one to diminish the bias of SLOPE.  However, there still remains the problem of the loss of FDR control, which typically occurs 
when the columns of the design matrix are correlated.
This loss of FDR control results from over-shrinkage of large regression coefficients, whose unexplained effect is often compensated by even slightly correlated ``false'' explanatory variables (see \citet{su2017false} for the theoretical analysis of the similar phenomenon for the LASSO).

\subsection{Our Contribution}
The adaptive Bayesian version of SLOPE (ABSLOPE) we propose here addresses these issues by incorporating aspects of the Spike-and-Slab  LASSO. By embedding SLOPE within a Bayesian spike-and-slab framework, our  prior is  constructed  so that the ``spike'' component effectively  reduces to  regular SLOPE   for very small regression coefficients.  Together with  a bias-reducing slab  for large signals,  this allows for FDR control under a wide range of possible scenarios, as will be seen from our extensive simulation study. 
In addition, the ``slab'' component of our mixture prior  preserves the averaging property of SLOPE for similar regression coefficients (see \citet{OWL} for discussion of the SLOPE averaging effect). This  leads to very good prediction properties when regressors are substantially correlated. 
The hyper-parameters of our mixture SLOPE prior are iteratively updated using the full Bayesian model in the spirit of stochastic approximation EM \citep{lavielle:hal-01122873}, which can also handle missing data. 

Our aim is to develop a complete and efficient methodology  for selection of variables with high dimensional data and missing values. 
The methodology  has been implemented in an R \citep{softR} package  {\sc ABSLOPE} \citep{ABSl1}. The code that reproduces all our experiments is available from GitHub \citep{github}.

\subsection{Previous work on selecting variables with missing data}
Handling missing data within the context of high-dimensional variable selection is a very important problem. Indeed, missing data are omnipresent. For example, genetic data obtained from microarray experiments often contain missing values for several reasons: insufficient resolution, image corruption, manufacturing errors, etc.  
The most common practice of dealing with missing data, i.e. listwise deletion, leads to estimation bias, unless the missing data are generated completely randomly, and  information loss. There is no shortage of literature on  missing values management, e.g. see \citet{little_rubin} and the platform \texttt{R-miss-tastic}\footnote{\url{https://rmisstastic.netlify.com}} \citep{rmiss} for an overview of the state of the art.
However, there are only a few methods for selecting an actual model when  covariate values are missing.
For example, in generalized linear models, \citet{missaic,icq,wei2018}  adapted likelihood-based information criteria designed for complete data such as AIC. However, their methods cannot process large data where the dimension $p$ is larger
than (or comparable to)  the sample size $n$. In linear models, \citet{loh2012} formulated a LASSO variant by modifying the covariance matrix estimation for the case of missing values, and solved the resulting non-convex problem with an algorithm based on the projected gradient descent.  However, this method assumes that the $l_1$ norm is bounded by a constant which depends on the sparsity level rarely known in practice. In other related work, \citet{zhao2017} suggested a  pseudo-likelihood method with a LASSO penalty, which can be used to select variables, but does not estimate the parameters. Finally, \citet{mirl}  combined penalized regression techniques with multiple imputation and stability selection.
}

This manuscript is organized as follows: Section \ref{sec:model} introduces   notation and assumptions about our ABSLOPE model. Section \ref{sec:fr} describes the stochastic approximation EM algorithm (and its simplified variant) for processing missing data.  Section \ref{sec:simu} evaluates the methodology with a comprehensive simulation study focusing on power, FDR and estimation bias. In Section \ref{sec:real}, we apply our approach to a medical dataset of trauma patients to develop  a model that predicts the rate of platelets using  (incomplete) medical information collected by the ambulance. 
Finally,  Section \ref{sec:concluded} concludes our work with a discussion.

{
\section{Statistical model and assumptions}\label{sec:model}
Let $y=(y_i, 1\leq i \leq n)$ be a vector of $n$ responses, centered such that $\bar y=\frac{1}{n} \sum_{i=1}^n y_i=0$;  and let $X= (X_{ij}, 1\leq i \leq n, 1\leq j \leq p )$ be a  design matrix of dimension $n\times p$ standardized so that each column has mean 0 and a unit $l_2$ norm, \textit{i.e.}  $\sum_{i=1}^n X_{ij} = 0$ and $
\sum_{i=1}^n X_{ij}^2 =1$ for $1\leq j \leq p.$ 
We consider the problem of estimating $\beta$ based on realizations $y$ from the linear regression model:
$$
y=X\beta+\varepsilon,
$$
where   $\beta = (\beta_j, 1\leq j \leq p)$ is the vector of regression coefficients of length $p$, for which we assume a sparse structure, and $\varepsilon$ is a  vector of length $n$ of independent Gaussian errors  with mean 0 and variance $\sigma^2$, \textit{i.e.} $\varepsilon \sim \nmdist(0,\sigma^2 I_n)$.

\subsection{SLOPE}\label{ssec:slope}
SLOPE \citep{bogdan2015} estimates coefficients by minimizing a regularized residual sum of squares using a sorted $l_1$ norm penalty which generalizes  the LASSO by penalizing larger coefficients more stringently: 
\begin{equation}\label{eq:slope}
\bslope=\argmin_{\beta \in \mathbb{R}^p}\left\{ \frac{1}{2} \Vert y- X\beta \Vert^2 + \sigma \sum_{j=1}^p \lambda_j|\beta|_{(j)}\right\}\;,
\end{equation}
where the penalty coefficients $\lambda_1 \geq \lambda_2 \geq \cdots \geq \lambda_p \geq 0$ and the absolute values of elements in $\beta$ are sorted in a decreasing order $|\beta|_{(1)} \geq |\beta|_{(2)} \geq \cdots \geq |\beta|_{(p)}$. The sorted $l_1$ penalty can also be written as: $$\text{pen}(\lambda) =  \sigma \sum_{j=1}^p \lambda_j|\beta|_{(j)}= \sigma \sum_{j=1}^p \lambda_{r(\beta,j)}|\beta_j| \; ,$$
where $r(\beta,j) \in \{1,2,\cdots,p\}$ is the rank of $\beta_j$  among elements in $\beta$ in a descending order. 
To solve the convex but non-smooth optimization problem (\ref{eq:slope}), a proximal gradient algorithm can be used as detailed in \citet{bogdan2015}.
Unlike in SSL, the SLOPE formulation operates under the following premise: the higher the rank (i.e. the stronger the signal), the larger the penalty. 
This behavior is quite similar to the Benjamini-Hochberg procedure (BH) \citep{bhq}, which compares more significant $p$-values with more stringent thresholds. In this way, SLOPE can be seen as building a bridge between the LASSO and the False Discovery Rate (FDR) control for multiple testing. 
In the context of multiple regression we define FDR of an estimator $\hat \beta=(\hat \beta_1,\ldots,\hat \beta_p)$ as
$$\rm{FDR}=\mathbb{E}\left(\frac{V}{\rm{max}(1,R)}\right)\;,$$
where $$R=\#\{j: \hat \beta_j\neq 0\}\;\;\mbox{and}\;\;V=\#\{j: \hat \beta_j\neq 0 \wedge \beta_j=0\}\;.$$
 SLOPE  \citep{bogdan2015} uses the sequence of parameters $\lambda_{\text{BH}}=(\lambda_{\text{BH},1},\ldots, \lambda_{\text{BH},p})$  with 
$$ \lambda_{\text{BH},j} = \Phi^{-1}\left(1-j \times \frac{q}{2p}\right)\;,$$
where $\Phi(\cdot)$ denotes the cdf of $\mathcal{N}(0,1)$ and $q$ is the target FDR level. \color{black}

\subsection{Adaptive Bayesian SLOPE}



As with any other penalized likelihood estimator, SLOPE can be seen as a posterior mode under the following prior \citep{amir2016}: 
\begin{equation*}\label{eq:slopeprior}
\dens(\beta \mid \sigma^2;\lambda)= C(\lambda,\sig2)\prod_{j=1}^p  \exp\left(-\frac{1}{\sigma}\lambda_{r(\beta,j)}|\beta_j| \right)\;,
\end{equation*}
where $C(\lambda,\sig2)$ is a normalizing constant.


This prior depends on just one sequence of tuning parameters $\lambda$, which regulates both  model selection and shrinkage. Simulation results reported in \citet{bogdan2015} show that the selection of $\lambda$ leading to FDR control also leads to over-excessive shrinkage and large estimation bias. To solve this problem we follow the idea of the Spike-and-Slab LASSO (SSL) \citep{spikeslablasso}.
SSL avoids over-shrinkage of large effects with a two-point   Laplace mixture prior, where large coefficients can escape shrinkage by migrating towards the slab portion of the prior. 
The spike component is assigned a large penalty $\lambda_0$ (small variance) to weed out noise, while the slab component has a small penalty $\lambda_1$ (large variance) to provide enough support for large signals.  The Spike-and-Slab LASSO procedure is based on   maximum a posteriori estimation (MAP) which relies on fast weighted LASSO calculations with  weights automatically  adjusted throughout the algorithm. Namely, separately for each variable we have a penalty which  depends on the (conditional) posterior probability that this variable is an important predictor. 
The SSL prior also automatically learns the level of sparsity through an empirical-Bayes plug-in inside the algorithm. 
The optimal choice of the spike penalty $\lambda_0$ relates to the prior mixing  weight $\theta$ and should  reflect the inherent sparsity of the signal \citep{ssl2}. The SSL procedure does not choose a single value $\lambda_0$ but, similarly as the LASSO, creates  a solution path indexed by increasing values of $\lambda_0$.
Since the SLOPE procedure was shown to be adaptive to the level of sparsity, we will replace the spike portion of the SSL prior with the Bayesian SLOPE prior to achieve more automatic sparsity adaptation.

 In our adaptive Bayesian SLOPE (ABSLOPE), we thereby consider a different hierarchical Bayesian model with the spike prior based on the sequence of SLOPE decaying parameters to provide FDR control and  with the SLOPE slab prior to stabilize estimation of large signals by additional shrinkage of regression parameters towards one another (see \citet{grpSLOPE} for some discussion of the SLOPE shrinkage). 
 ABSLOPE   borrows strength across covariates (by tying them together through the spike distribution)
 and, similarly as SSL, allows for  estimation of latent inclusion parameters and the level of sparsity (i.e. number of nonzero   $\beta$ coefficients). 
The procedure requires only three interpretable input parameters: FDR level $q$ and the hyperparameters $a$ and $b$ of the Beta prior for the sparsity level $\theta\sim Beta(a,b)$.

The ABSLOPE prior on  the regression vector $\beta$ is formally defined as:  \begin{equation}\label{eq:priorbeta} 
\dens(\beta\mid \gamma,c,\sigma^2;\lambda) \propto c^{\sum_{j=1}^p \iden(\gamma_j=1)}\prod_{j=1}^p \exp\left\{-w_j \lvert\beta_j\rvert\frac{1}{\sigma} \lambda_{r(W\beta,j)} \right\}.
\end{equation}
This formulation may seem a bit complicated  at first sight and so we carefully explain its components below:
\begin{enumerate}
\item Each $\beta_j\neq 0$ is regarded as signal and noise otherwise.
\item As is customary with spike-and-slab priors, each covariate $x_j$ is equipped with a binary inclusion indicator $\gamma_j \in \{ 0,1\}$ which indicates whether $\beta_j$ is 
is substantially different from the noise level. The vector $\gamma = (\gamma_1, \cdots, \gamma_p)$ then indexes $2^p$ possible model configurations. Conditionally on a mixing (prior inclusion) weight $\theta\in(0,1)$, we define the model distribution as an independent Bernoulli product:
$$\dens(\gamma\mid \theta)=\prod_{j=1}^p \theta^{\gamma_j}(1-\theta)^{1-\gamma_j}\;,$$
where 
$\theta = \mathbb{P}(\gamma_j=1; \theta)$ is formally defined as the expected fraction of large $\beta_j$, \textit{i.e.}, $\theta$ indicates the level of sparsity. We assume that $\theta$ arose from a beta distribution $Beta(a,b)$, where the values of $a$ and $b$ can be selected by the user, according to an initial guess of the signal sparsity.
\item  The parameter
$c \in (0,1)$  is the ratio of average signal magnitudes between the null components and the non-null components. We assume a non-informative prior $c \sim \mathcal{U}[0,1].$
\item We define a diagonal weighting matrix $W = \text{diag}(w_1, w_2, \cdots, w_p)$  consisting of  elements $$w_j=c\gamma_j+(1-\gamma_j)=\begin{cases} c, & \gamma_j =1\\
1, & \gamma_j=0 \end{cases}\;.$$ 
\item  For the case when the noise variance $\sigma$ is unknown, we assume an uninformative prior  $\dens(\sigma^2)\propto \frac{1}{\sigma^2}$. 
\end{enumerate}

\subsection{Motivation}
In Appendix \ref{ann:dev}  it is proved that the prior (\ref{eq:priorbeta}) leads to the regular SLOPE prior on the transformed parameter
vector $z=W\beta$, i.e.
\begin{equation}\label{eq:zslope}
\dens(z \mid \sigma^2;\lambda) \propto \prod_{j=1}^p  \exp\left\{-\frac{1}{\sigma}\lambda_{r(z,j)}|z_j| \right\}\;,
\end{equation}

As a result, when $W$ is known (i.e. we know the signal and noise variables from $\gamma_j\in\{0,1\}$) and when the data are fully observed, the MAP for $\beta$ under the ABSLOPE prior (\ref{eq:priorbeta})  can be obtained as a  solution to  SLOPE (\ref{eq:slope}) with a weighted design matrix $\tilde{X} = XW^{-1}$. 
Let us now clarify the value of introducing the weighting matrix $W$. It turns out that when $\gamma_j=0$ we have $w_j = 1$, \textit{i.e.},  noise variables are treated with the regular SLOPE penalty which 
will assign substantially larger shrinkage to smaller effects. This is different from the SSL prior, which would shrink all the noise coefficients equally by $\lambda_0$.
On the other hand, when $\gamma_j=1$ we have $w_j = c<1$ and the variables are treated as  true signals  and thereby not shrunk as much. 
This is achieved by multiplying the respective elements of the vector of tuning parameters by $c$ and, additionally, by moving these variables towards the end of sequence. 
This implies that, under ABSLOPE, the large effects  $\beta_j$ will be assigned a  penalty $c\lambda_{r(W\beta,j)}$ \color{black} that is smaller than  $\lambda_{r(\beta,j)}$ obtained under the  regular SLOPE. As a result,  this adaptive version is poised to yield more accurate estimation since the  $l_1$ penalty on true signals will be much smaller.

\begin{figure}[!htbp]
\centering
\subfloat[Non-null $\beta$]{\includegraphics[width=0.45\textwidth]{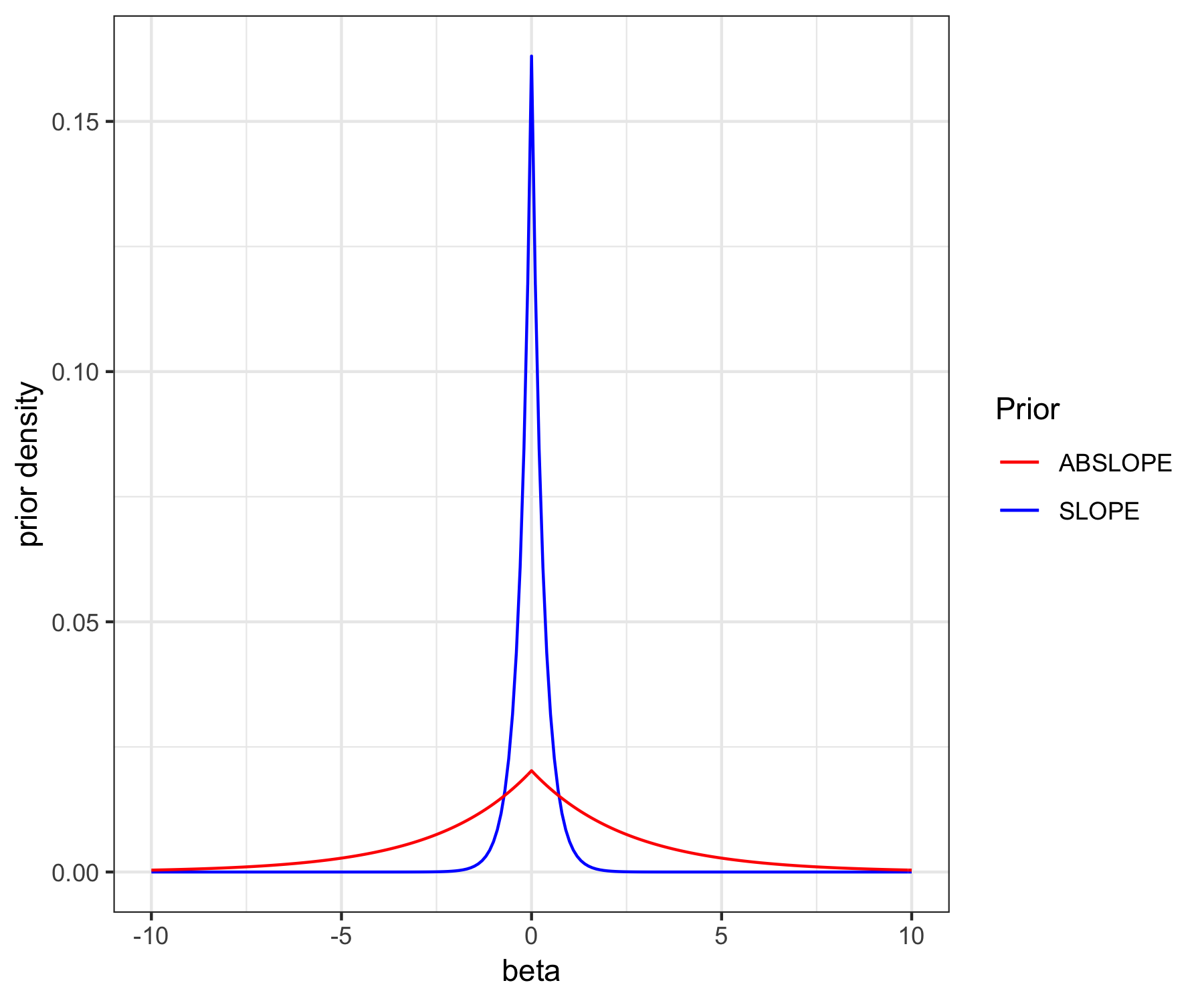}}
\hfill
\subfloat[Null $\beta$]{\includegraphics[width=0.45\textwidth]{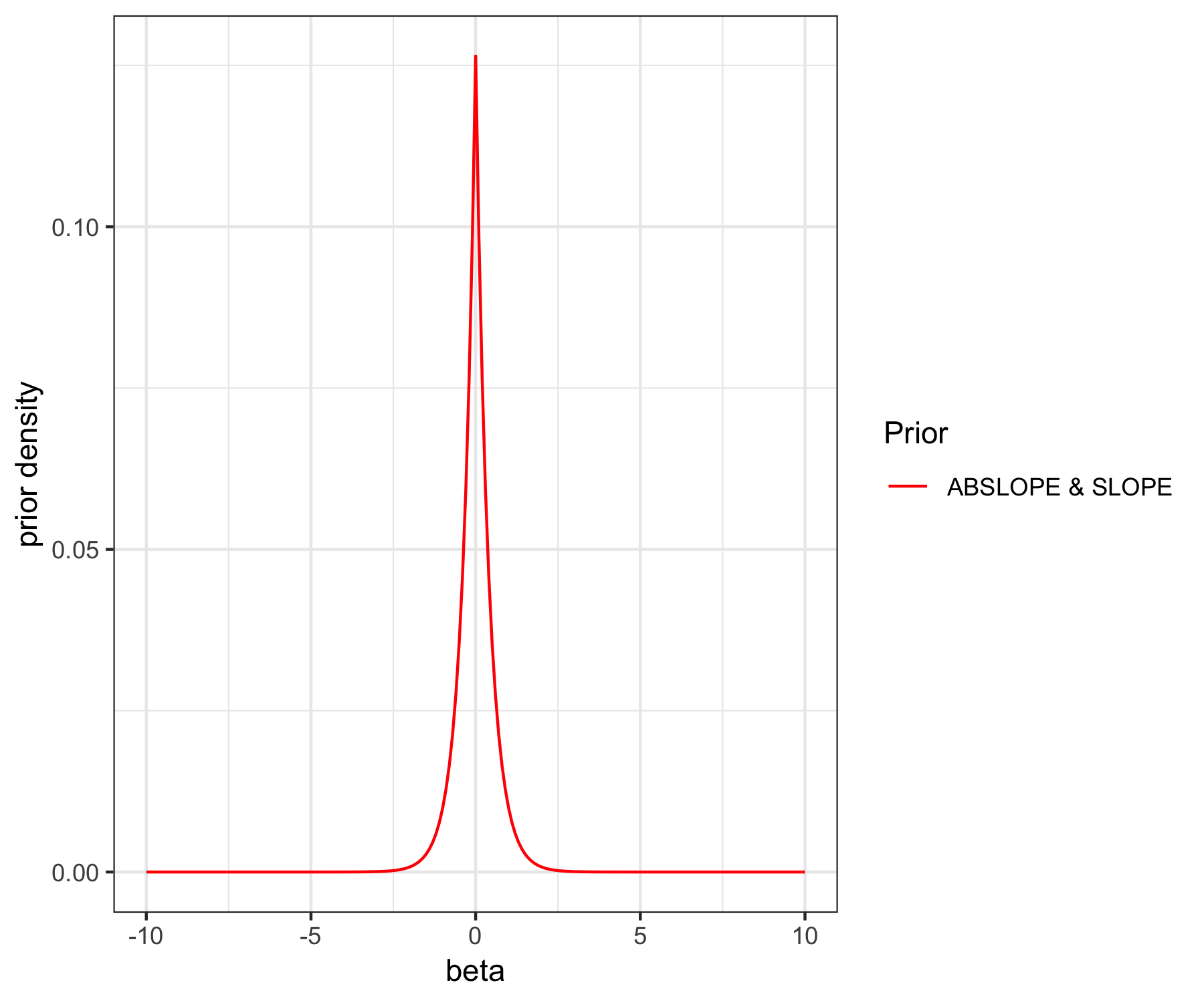}}
\caption{Prior distribution of SLOPE and ABSLOPE, on $\beta$ whose true value is non-null (a) or null (b).}
\label{fig:prior}
\end{figure}
 Figure \ref{fig:prior} shows the difference between the SLOPE prior and the ABLSOPE prior on a single coefficient $\beta_j$.
 On the left, we have a  slab prior   distribution on an active coefficient $\beta_j$  which shows that ABSLOPE  promotes larger estimates: the mass is greater in the tails compared to SLOPE. On the other hand, 
 for the irrelevant $\beta_j$ (spike prior depicted on the right), ABSLOPE  reduces to the double exponential SLOPE peak to threshold out small effects. 

The ABSLOPE prior can be seen as a spike-and-slab prior, where  the spike component models regression coefficients close to the noise level and the slab component models large regression coefficients. 
In fact, the spike-and-slab LASSO prior can be regarded as a special case when one considers 
 the constant sequence of tuning parameters  $\lambda_1=\ldots=\lambda_p=\lambda_0$ for the spike SLOPE component  and   $c$ as  the ratio between spike and slab penalties.
The algorithm described in Section \ref{ssec:slob} shows that the slab component is destined to de-bias the large regression coefficients while the spike component is aimed at  FDR control. 


\subsection{Assumptions for missing values} 
We suppose that the missingness occurs only in the covariates $X$, not in the responses $y$. For each individual $i$, we denote  with $\xiobs$ the observed elements of $\bx_{i} = (X_{i1}, X_{i2}, \cdots, X_{ip})$ and with $\ximis$ the missing ones. We also decompose the matrix of covariates  as $\bx = (\xobs,\xmis)$, keeping in mind that the missing elements may differ from one individual to another. For each individual $i$, we define the missing data indicator vector $m_i=(m_{ij}, 1 \leq j \leq p)$, with $m_{ij}=1$ if $X_{ij}$ is missing and $m_{ij}=0$ otherwise. The matrix $m=(m_i, 1\leq i \leq n)$ then defines the missing data pattern. The missing data mechanism is characterized by the conditional distribution of $m$ given $X$ and $y$, with a parameter $\phi$, \textit{i.e.},
$\dens(m_i\mid\bx_i,y_i,\phi).$ 
In the literature on missing data \citep{little_rubin}, three mechanisms \citep{rubin1966mechanism} are recognized to describe the distribution/sources of  missingness:  \emph{i)} Missing completely at random (MCAR): the absence is not related to any variable in the study; \emph{ii)} Missing at random (MAR): the missing data depends only on the observed variables; \emph{iii)}  Missing not at random (MNAR): the absence depends on the value itself.
Throughout this paper, we assume the MAR mechanism which implies that the missing values mechanism  can therefore be ignored when maximizing the likelihood \citep{little_rubin}. A reminder of these concepts is given in the Appendix \ref{ann:ignorable}.
 
 We adopt a probabilistic framework by assuming that 
 $\bx_i = (X_{i1},\ldots, X_{ip})$ is normally distributed:
\begin{equation*}
\bx_i \iid \mathcal{N}_p(\mu,\Sigma), \quad i=1,\cdots,n \;. 
\end{equation*}

Since the covariates should be standardized (as we assumed at the beginning of Section \ref{sec:model}), we have to reconsider our scaling of $X$ in the light of missing data. When the missing values are MCAR,  scaling can be performed as a pre-processing step before performing the analysis. Since observed values represent a random sample from the population, standard deviations estimated using observed data   are unbiased estimates of the population standard deviation even if their variance is larger. 
When the missing data are MAR, standard deviations estimated using observed data can be severely biased. Indeed, consider the case when two variables are highly correlated and missing values occur in one variable when the values of the other variable are larger than a constant, then the estimated standard deviation will be biased downwards. Consequently, its estimation needs to be included in the analysis. In the Appendix \ref{ann:std}, we detail how we update mean and standard deviation at each iteration of the algorithm presented in Section \ref{sec:fr}.

\subsection{Overview of modeling}
Figure \ref{fig:gm} shows our ABSLOPE  graphical model with  variables, parameters and their relations. We aim at estimating $\beta$ and $\sigma^2$, treating parameters $\mu$ and $\Sigma$
as nuisance.
\begin{figure}[!htb]
\centering
\includegraphics[width=0.6\textwidth]{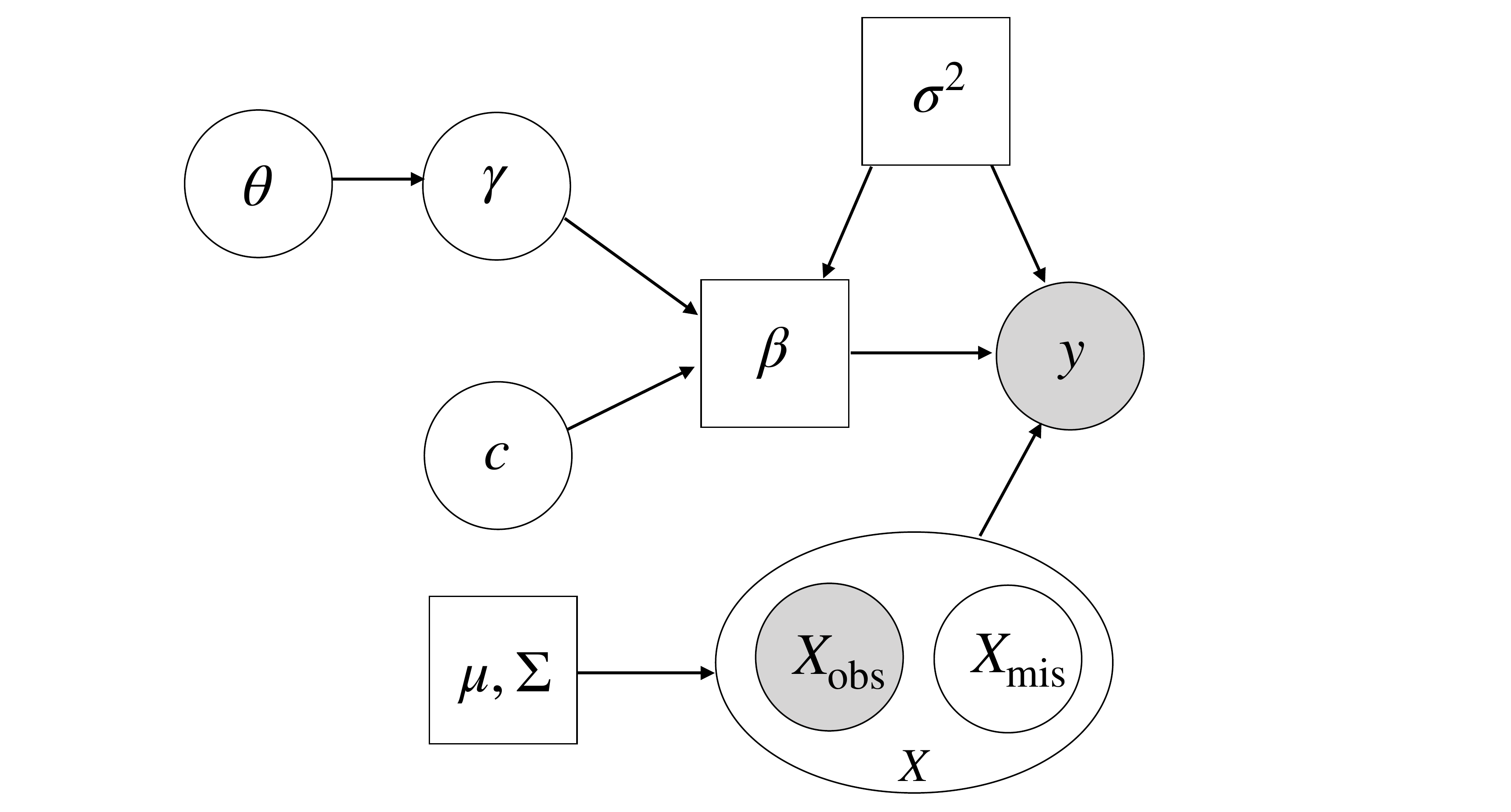}
\caption{ABSLOPE graphical model. Arrows indicate dependencies. White circles are for latent variables, gray ones for observed variables and squares for parameters.}
\label{fig:gm}
\end{figure}

\section{Parameter estimation and model selection}\label{sec:fr}
In this section,  we develop an  ABSLOPE  method based on the stochastic approximation EM algorithm. 
As this algorithm entails proper sampling which can be quite time consuming, we also provide a  simplified heuristic version called SLOBE, where the stochastic step is replaced with  deterministic approximations of parameter expected values. This faster variant allows us to consider models of larger dimensions and, according to our simulation study, performs very similarly to the stochastic version.

\subsection{Maximizing the observed penalized likelihood}
According to the model defined in Section \ref{sec:model} and presented in Figure \ref{fig:gm},  the penalized complete-data log-likelihood can be written as:
\begin{equation}\label{eq:llcomp}
\begin{split}
\ell_{\rm comp} = &\log \dens(y,X,\gamma,c;\,\beta,\theta,\sigma^2) + pen(\beta)\\
=&\log \left\{ \dens(X\mid  \mu,\Sigma)\,\dens(y\mid X;\,\beta, \sigma^2)\,\dens(\gamma \mid \theta)\,\dens(c) \right\}+ pen(\beta)\\
= & -\frac{1}{2} \log ( 2\pi \lvert \Sigma \rvert) -\frac{1}{2} (X-\mu)^T\Sigma^{-1}(X-\mu )
-n\log(\sigma) -  \frac{1}{2\sigma^2}\Vert y-X\beta \Vert^2  \\
&+ \sum_{j=1}^p \iden(\gamma_j=1)\log\theta  + \sum_{j=1}^p \iden(\gamma_j=0)\log(1-\theta) 
- \frac{1}{\sigma} \sum_{j=1}^p w_j\lvert\beta_j\rvert{\lambda_{r(W\beta,j)}} .
\end{split}
\end{equation}
Similarly as the EMVS variable selection procedure of \cite{EMVS}, we focus on obtaining the  MAP  point estimates and do not aspire at  fully Bayesian inference which would entail calculating the entire posterior distribution.
Due to the presence of latent variables $\xmis, \gamma$ and $c$, we estimate $\beta$ by maximizing the observed log-likelihood which integrates over the latent variables:
$\llike_{\rm obs} = \iiint \llike_{\rm comp} \, d\xmis \, dc \, d\gamma.$ We use the EM algorithm \citep{em1977} to  estimate $\beta$, and in the meantime, obtain  simulated $\gamma$ to distinguish the true signals from the noise, \textit{i.e.} to  select variables. Given the initialization, each iteration $t$ updates $\beta^t$ to $\beta^{t+1}$ with the following two steps:
\begin{itemize}
\item \emph{E step: }
The expectation of the complete-data log likelihood with respect to the conditional distribution of latent variables is computed, \textit{i.e.}, $$Q^t = \mathbb{E}(\llike_{\rm comp}) \quad \text{wrt} \quad \dens(\xmis,\gamma,c, \theta\mid y,\xobs,\beta^{t},\sigma^{t},\mu^{t},\Sigma^{t})\;.$$
Since this is not tractable, we derive a stochastic approximation EM (SAEM) algorithm \citep{lavielle:hal-01122873} by replacing the  E step by a simulation step and a stochastic approximation step.
\begin{itemize}
\item \emph{Simulation:} draw one sample $(\xmis^{t},\gamma^{t},c^{t},\theta^{t})$ from 
\begin{equation}\label{eq:simusaem}
\dens(\xmis,\gamma,c, \theta\mid y,\xobs,\beta^{t-1},\sigma^{t-1},\mu^{t-1},\Sigma^{t-1})\;;
\end{equation}
\item \emph{Stochastic approximation:} update function Q with 
\begin{equation}\label{eq:appsaem}Q^{t} = Q^{t-1} + \eta_t \left(\llike_{\rm comp} {\Big |}_{\xmis^{t},\gamma^{t},c^{t},\theta^{t}}-Q^{t-1}\right)\;,
\end{equation}
where $\eta_t$ is the step-size.
\end{itemize}
The step-size $(\eta_t )$ is chosen as a decreasing sequence as described in \citet{convergence_saem} which ensures almost sure convergence of SAEM to a maximum of the observed likelihood in their continuously differentiable case. 
\item \emph{M step:}
$(\beta^{t+1},\sigma^{t+1},\mu^{t+1},\Sigma^{t+1})=\argmax Q^{t+1}.$\\
Note that $\Sigma^{t+1}$ is estimated as above only when $p <<n$. Otherwise we consider a shrinkage estimation as discussed in Remark \ref{rm:shrinkvar}. Indeed, we regard $(\mu, \Sigma)$ as auxiliary parameters, which are needed only to update the missing values.
\end{itemize}
Despite the apparent complexity of the algorithm, it turns out that the likelihood (\ref{eq:llcomp}) can be decomposed into several terms: one term for the linear regression part, one term for the covariates distribution and terms for the latent variables $\gamma$ and $c$, as illustrated in Figure \ref{fig:gm}. Consequently, one iteration can be divided into tractable sub-problems, as detailed in the following subsections.
\subsection{Simulation step: sampling the latent variables}\label{ssec:simu}
To perform the simulation step \eqref{eq:simusaem}, we use the Gibbs sampler. To simplify notation, we hide the superscript and note that all  conditional distributions are computed  given the quantities from the previous iteration. 
We perform the following sampling procedure:
\begin{equation}\label{eq:simu}
\begin{cases}\, \gamma \sim Bin\left(\frac
{\theta c \exp\left(-c{\frac{1}{\sigma}}\lvert\beta_j\rvert\lambda_{r(W\beta,j)}\right)}{(1-\theta)\exp\left(-{\frac{1}{\sigma}}\lvert\beta_j\rvert\lambda_{r(W\beta,j)}\right)  + \theta c \exp\left(-c{\frac{1}{\sigma}}\lvert\beta_j\rvert\lambda_{r(W\beta,j)}\right)} \right) \,; \\ 
\, \theta \sim Beta \left(a+\sum_{j=1}^p \iden (\gamma_j=1), b+\sum_{j=1}^p \iden(\gamma_j=0)\right), \text{ with } Beta(a,b) \text{ a prior for } \theta  \,;\\ 
\, c\sim Gamma \left(1+\sum_{j=1}^p\iden(\gamma_j=1), \quad \frac{1}{\sigma}\sum_{j=1}^p \lvert\beta_j\rvert \lambda_{r(W\beta,j)}\iden(\gamma_j=1) \right) \text{ truncated to } [0,1]. \end{cases}
\end{equation}
The detailed calculation and  interpretation can be found in Appendix \ref{ann:simu}.
In addition, to simulate the missing values $\xmis$, we perform a decomposition:
\begin{equation}\label{eq:xmis}
\begin{split}
\xmis &\sim \dens(\xmis\mid \gamma,c,y,\xobs,\beta,\sigma,\theta,\mu,\Sigma)\\
 & =\dens(\xmis\mid y,\xobs,\beta,\sigma,\mu,\Sigma)\\
&\propto \dens(y\mid \xobs,\xmis,\beta,\sigma )\, \dens(\xmis\mid\xobs,\mu,\Sigma)\;.
\end{split}
\end{equation}
Here, we observe that the target distribution (\ref{eq:xmis}) is a normal distribution since the two terms after factorization are both normal. In the following proposition, we give the explicit form of the target distribution as a solution to  a system of linear equations.
\begin{prop}\label{prop:xmislineq}
For a single observation $x=(\xm,\xo)$  we denote with $\xo$ and $\xm$   observed and missing covariates, respectively.  Let $ \mathcal{M}$ be the set containing indexes for missing covariates and $\mathcal{O}$ for the observed ones.
Assume that $p(\xo,\xm; \Sigma, \mu)\sim \mathcal{N}(\mu,\Sigma)$ and let $y=x\beta+\varepsilon$ where $\varepsilon\sim N(0,\sigma^2)$.
For all the indexes of the missing covariates $ i \in \mathcal{M}$, we denote: 
\[
 m_i = \sum_{q=1}^p \mu_j s_{iq},\quad  u_i = \sum_{k \in \mathcal{O}}\xo^k s_{ik},\quad r = y-\xo{\beta_{\rm obs}},\quad \tau_i =\sqrt{s_{ii}+\beta_i^2/\sigma^2}\;,
\]
with $s_{ij}$ elements of $\Sigma^{-1}$ and ${\beta_{\rm obs}}$ the observed elements of $\beta$.\\
Let $\tilde\mu=(\tilde \mu_i)_{i\in\mathcal{M}}$ be the solution of the following system of  linear equations: 
\begin{equation}\label{eq:xmislineq}
 \frac{
r\beta_i /\sigma^2
+m_i-u_i}{\tau_i} 
-
   \sum_{j \in \mathcal{M}, j\neq i}\frac{
\beta_i\beta_j
  /\sigma^2+s_{ij}}{\tau_i\tau_j}\tilde\mu_j = \tilde \mu_i\;, \quad \text{for all }
 i \in \mathcal{M}\;,
\end{equation}
and let $B$ be a matrix with elements: 
\[
 B_{ij}= \begin{cases}  \frac{
 \beta_i\beta_j
 /\sigma^2+s_{ij}}{\tau_i\tau_j}, & \text{if  } i \neq j \\
  1, & \text{if  } i=j 
   \end{cases}\;,
\]
then for $z=(z_i)_{i \in \mathcal{M}}$ where $z_i = \tau_i \xm^i$ we have:
\[
 z \mid \xo, y; \Sigma,\mu,\beta,\sigma^2 \sim N(\tilde\mu, B^{-1})\;.
\]
As a result, we can simulate missing covariates from:\[
 \xm \mid \xo, y; \Sigma,\mu,\beta,\sigma^2 \sim N(\tilde\mu \oslash \tau, B^{-1} \oslash (\tau \tau^T))\;,
\]
where $\tau = (\tau_i)_{i \in \mathcal{M}}$ $\oslash$ is used for Hadamard division. 
The proof is provided in Appendix \ref{ann:xmislineq}.
\end{prop}

\subsection{Stochastic approximation and maximization steps}\label{ssec:app}
After the simulation step, we obtain one sample for each latent variable: $\xmis^t, \gamma^t$, $c^t$, and thus $W^t$  with diagonal elements $w^t_j = 1-(1-c^t)\gamma_{j}^t$.
Now we have several parameters to estimate, but each parameter only concerns some of the terms in the complete-data likelihood. This  helps us  simplify calculations. 
The maximization step is nevertheless quite difficult because the complete model does not belong to a regular exponential family (if so we could update the sufficient statistics and  maximize more easily).

As the implementation of   SAEM is quite challenging in the general step-size case, we start with the simpler case of fixed step-size $\eta_t=1$. It is important to note that this  causes larger variance compared to setting the step-size as a decreasing sequence \citep{convergence_saem} and there is no guarantee of convergence to the actual mode,  only to its neighborhood.
\subsubsection{Step-size $\eta_t=1$}\label{ssc:eta1}
When   $\eta_t=1$,   estimation  boils down to maximizing the complete-data likelihood completed by sampling the latent variables from their conditional distribution given the observed values .
\begin{enumerate}
\item Update $\beta$.
\begin{equation*}
\begin{split}
\beta^t = \argmax_{\beta}Q_1^{t}(\beta) := -  \frac{1}{2{(\sigma^{t-1})}^2}\Vert y-X^{t}\beta\Vert^2 - \frac{1}{\sigma^{t-1}} \sum_{j=1}^p w^{t}_j \lvert{\beta_j}\rvert {\lambda_{r(W^{t}\beta,j)}}  \,,
\end{split}
\end{equation*}
where $X^{t} = (\xobs, \xmis^{t})$.
This estimate corresponds to the solution of SLOPE, given the value of $W$, $\xmis$ and $\sigma$.  In our implementation of ABSLOPE we solve the SLOPE optimization problem using the Alternative Direction Method of Multipliers of \citep{ADMM}, which turns out to be much quicker then the proximal gradient algorithm of \citep{bogdan2015} when the regressors are strongly correlated or when they are on different scales, as in our reweighting scheme.

\item Update $\sigma$.
\begin{equation*}
\begin{split}
\sigma^{t}= \argmax_{\sigma}Q_2^{t}(\sigma) := 
-n\log(\sigma) -  \frac{1}{2\sigma^2}\lVert y-X^{t}\beta^{t} \rVert^2 - \frac{1}{\sigma} \sum_{j=1}^p w^{t}_j \lvert\beta^{t}_j\rvert {\lambda_{r(W^{t}\beta^{t},j)}} \;.
\end{split}
\end{equation*}
Given by the derivative, the solution to estimate $\sigma$ is:
\begin{equation}\label{eq:sigma}
\sigma^{t} = \frac{1}{2n}\left[ \sum_{j=1}^p \lambda_{r(W^{t}\beta^{t},j) } w^{t}_j \lvert \beta^{t}_j \rvert + \sqrt{\left( \sum_{j=1}^p \lambda_{r(W^{t}\beta^{t},j) } w^{t}_j \lvert \beta^{t}_j \rvert \right)^2 + 4n\rm{RSS}}  \right]\;,
\end{equation}
where the RSS (residual sum of squares) is $\lVert y-X^{t}\beta^{t} \rVert^2 $.

If we omit the penalization term, (\ref{eq:sigma}) amounts to $\sigma^t  =\sqrt{\frac{RSS}{n}}$, which is the classical formula for MLE of $\sigma$ when $\beta$ is also estimated by MLE. In this case this estimator would be biased downwards. Interestingly, our posterior mode estimator of $\sqrt{n}\sigma$ is larger than the corresponding RSS, which, according to the simulation results in Subsection \ref{ssec:covg}, often leads to a less biased estimator when most of the true effects are detected by ABSLOPE.


\item Update $\mu, \Sigma$:
$$\mu^t, \Sigma^t = \argmax_{\mu, \Sigma}-\frac{1}{2} \log ( 2\pi \lvert\Sigma\rvert) -\frac{1}{2} (X^t-\mu)^{\top}\Sigma^{-1}(X^t-\mu )\;.$$
When $p<<n$, the solution is given by the empirical mean and  the empirical covariance matrix:
\begin{equation*}
\begin{split}
\mu^{t} = \bar{X}^{t}=\frac{1}{n}\sum_{i=1}^n X_i^{t}
\quad	  \text{and} \quad \Sigma^{t} = \frac{1}{n}\sum_{i=1}^n (X_i^{t} - \bar{X}^{t})(X_i^{t} - \bar{X}^{t})^{\top}\;.
\end{split}
\end{equation*}
In high dimensional setting,  estimation of $\Sigma^{t}$ by the empirical covariance matrix is replaced by  shrinkage estimation, as discussed in Remark \ref{rm:shrinkvar}.

\end{enumerate}
\begin{remark}\label{rm:shrinkvar}
To tackle the problem of estimation and inversion of the covariance matrix in high dimensions, 
one can resort to  shrinkage estimation as detailed in  \citet{ledoit2004}.
With the assumption that the ratio $\frac{n}{p}$ is bounded, they propose an optimal linear shrinkage estimator as a linear combination of identity matrix $I_p$ and the empirical covariance matrix $S$, i.e.: $$\hat{\Sigma} = \rho_1 I_p + \rho_2 S, \qquad \text{where } \rho_1, \rho_2 = \argmin_{\rho_1, \rho_2} \mathbb{E} \Vert \hat{\Sigma} - \Sigma \Vert^2\;.$$ The method boils down to shrinking  empirical eigenvalues towards their mean.
The parameters $\rho_1$ and $\rho_2$ are chosen with asymptotically (as $n$ and $p$ go to infinity) uniformly minimum quadratic risk in its class. 
\end{remark}

\subsubsection{General step-size}
With a general step-size  (say $\eta_t = \frac{1}{t}$), for a model parameter $\psi$   
we set
\begin{equation}\label{eq:prop}
\psi^{t+1} = \psi^{t} + \eta_t \left[ \hat{\psi}^t_{MLE} -\psi^{t} \right]\;, 
\end{equation}
where $\hat{\psi}^t_{MLE}$ is the MLE estimator of the complete-data likelihood completed by drawing the latent variables from their conditional distributions given the observed information. This exactly corresponds to the estimate  in Subsection \ref{ssc:eta1} when $\eta_t = 1$. 
In other words,  we apply stochastic approximations on the model parameters, instead of directly operating on the likelihood in (\ref{eq:appsaem}).
When the likelihood (\ref{eq:llcomp}) is a linear function of the parameters, the stochastic approximation step in equation (\ref{eq:appsaem}) corresponds exactly to our proposal (\ref{eq:prop}). 
In other situations, it gives good results from an empirical point of view.

\subsection{{SLOBE}: Quick version of {ABSLOPE}}\label{ssec:slob}
The implementation of SAEM, as described in  Subsection \ref{ssec:simu} and \ref{ssec:app}, can still be costly in terms of computation time, even if the terms of the  likelihood decompose well and we use the approximation (\ref{eq:prop}). We therefore propose  a simplified version of the algorithm, called {SLOBE}, which instead of drawing samples  $(\xmis^{t},\gamma^{t},c^{t}, \theta^{t})$ from their conditional distribution (\ref{eq:simusaem}) in the simulation step,   approximates them by  their conditional expectation, i.e.,
\begin{equation*}
(\xmis^{t},\gamma^{t},c^{t}, \theta^{t}) \leftarrow \mathbb{E}(\xmis,\gamma,c\mid y,\xobs,\beta^{t-1},\sigma^{t-1},\mu^{t-1},\Sigma^{t-1})\;;
\end{equation*}
To simplify  notation, we hide the superscript, but note that all the conditional expectations are computed given the quantities from the previous iteration. 
\begin{enumerate}
\item Approximate $\gamma_j$ by: 
\begin{equation}\label{eq:slobgamma}
\begin{split}
\pi &\coloneqq \mathbb{E}(\gamma_j=1\mid\gamma_{-j},c,\beta,\sigma,\theta, W) =  p( \gamma_j=1\mid\gamma_{-j},c,\beta,\sigma,\theta, W)\\
& \stackrel{\text{(\ref{eq:simu})}}{=}
\frac
{\theta c \exp\left(-c{\frac{1}{\sigma}}\lvert\beta_j\rvert\lambda_{r(W\beta,j)}\right)} {(1-\theta)\exp\left(-{\frac{1}{\sigma}}\lvert\beta_j\rvert\lambda_{r(W\beta,j)}\right)  + \theta c \exp\left(-c{\frac{1}{\sigma}}\lvert\beta_j\rvert\lambda_{r(W\beta,j)}\right) }\;.
\end{split}
\end{equation}
\item Approximate $\theta$ by:  \begin{equation}\label{eq:slobtheta}
\begin{split}
\mathbb{E}(\theta\mid\gamma,y,\xobs,\xmis,\beta,\sigma,c,\mu,\Sigma, W) = \mathbb{E}(\theta\mid\gamma,\beta,\sigma, W)
\stackrel{\text{(\ref{eq:simu})}}{=}  \frac{
a+\sum_{j=1}^p \iden (\gamma_j=1)}{a+ b+p}\;,
\end{split}
\end{equation}
where $a$ and $b$ are fixed parameters in the prior of $\theta$.
\item Approximate $c$ by:  \begin{equation}\label{eq:slobc}
\begin{split}
\mathbb{E}(c\mid\gamma,y,\xobs,\xmis,\beta,\sigma,\theta,\mu,\Sigma, W) \stackrel{\text{(\ref{eq:c})}}{=} 
\frac{\int_0^1 x^{a'}\exp(-b'x) dx}{\int_0^1 x^{a'-1}\exp(-b'x) dx}\;,
\end{split}
\end{equation}
where $a'= 1+\sum_{j=1}^p\iden(\gamma_j=1)$, $b'= \frac{1}{\sigma}\sum_{j=1}^p \lvert\beta_j\rvert \lambda_{r(W\beta,j)}\iden(\gamma_j=1) $.
\item In the case with missing values, for the $i^{\rm th}$ observation $X_i$, approximate $\ximis$ by:
\begin{equation*}
\mathbb{E}(\ximis\mid \gamma,c,y,\xiobs,\beta,\sigma,\theta,\mu,\Sigma) =\mathbb{E}(\ximis\mid y,\xiobs,\beta,\sigma,\mu,\Sigma)\;,
\end{equation*}
which is provided by Proposition \ref{prop:xmislineq}. 
\end{enumerate}
Then, in step M, we maximize the likelihood of the complete data, as in Subsection \ref{ssc:eta1}. 
The impact of replacing the simulation step with a conditional expectation is that we ignore the variability of latent variable sampling, which in  high dimensional settings helps reduce noise of the algorithm, and which also leads to  accelerations  as shown in  our simulation study in Subsection \ref{ssec:time}. 
We provide a summary of {ABSLOPE} and {SLOBE} methods in Appendix \ref{ann:algo}.

\section{Simulation study}\label{sec:simu}
\subsection{Simulation setting}

To illustrate the performance of our methodology, we perform simulations by first generating data sets as follows:
\begin{enumerate}
\item A  design matrix $X_{n \times p}$ is generated from a multivariate normal distribution  $\nmdist(\mu, \Sigma)\;$. The matrix is standardized, s.t., the mean of each column is 0 and its $\ell_2$-norm  is 1.
\item The signal magnitude is $c_0\sqrt{2\log p}$\footnote{This signal strength is inspired by the penalty coefficient of the Bonferroni method to control the family wise error rate (FWER) : $\lambda_{Bonf} = \sigma \phi^{-1}(1-\frac{\alpha}{2p}) \approx \sqrt{2\log p}$, for $p$ large and $\alpha$ fixed, say $\alpha=0.05$. } when $c_0$ is large the signal strength is stronger. Only $k$ on the $p$ predictors are non-zero and all equal to $c_0\sqrt{2\log p}$.  
\item The response vector is generated from $y = X\beta + \epsilon$ with  $\epsilon \sim N(0,\sigma^2 I_n)$ and $\sigma =1 $ to start.
\item Missing values are entered into the design matrix using a MCAR or MAR mechanism. 
For the former, we randomly generate 10\% of missing cells; for the later, we follow the multivariate imputation procedure proposed by \citet{ampute}. 
\end{enumerate}

We set the initialization and the hyperparameters as follows.
\paragraph{Initialization}
Appendix \ref{ann:init} provides the default values we have taken for the following simulation studies. The algorithm is not sensitive to the choice of values $a$ and $b$ (\ref{eq:slobgamma}), but initial values for $\beta$ may have a stronger impact. In practice, we use the LASSO estimates {based on preliminary mean imputation (missing values replaced by the average of the observed values for each variable) to initialize the coefficients.
\paragraph{Step-size} We set $\eta_t = 1$ for the first $t_0=20$ iterations to approach the neighborhood of the MLE, then, choose a positive decreasing sequence $\eta_t = \frac{1}{t - t_0}$ to approximate the MLE,  with the stochastic approach formula (\ref{eq:prop}).
\paragraph{$\lambda$ sequence}A sequence of penalty coefficients $\lambda$ must be chosen before implementing the algorithm. As introduced in the Subsection \ref{ssec:slope}, we use a  BH sequence inspired by orthogonal designs: $$\lambda_{BH}(j) = \phi^{-1}(1-q_j), \quad q_j =\frac{jq}{2p}, \quad j=1,2,\cdots,p.$$

\subsection{Convergence of SAEM} \label{ssec:covg}
We first illustrate the convergence of SAEM. We set the size of design matrix as $n=p=100$ while the number of true predictors is $k=10$, the signal strength $3\sqrt{2\log p}$ and the percentage of missingness $10\%$. 
The covariance $\Sigma$ is an identity matrix to start. 

\begin{figure}[!htbp]
\centering
\includegraphics[width=1\textwidth]{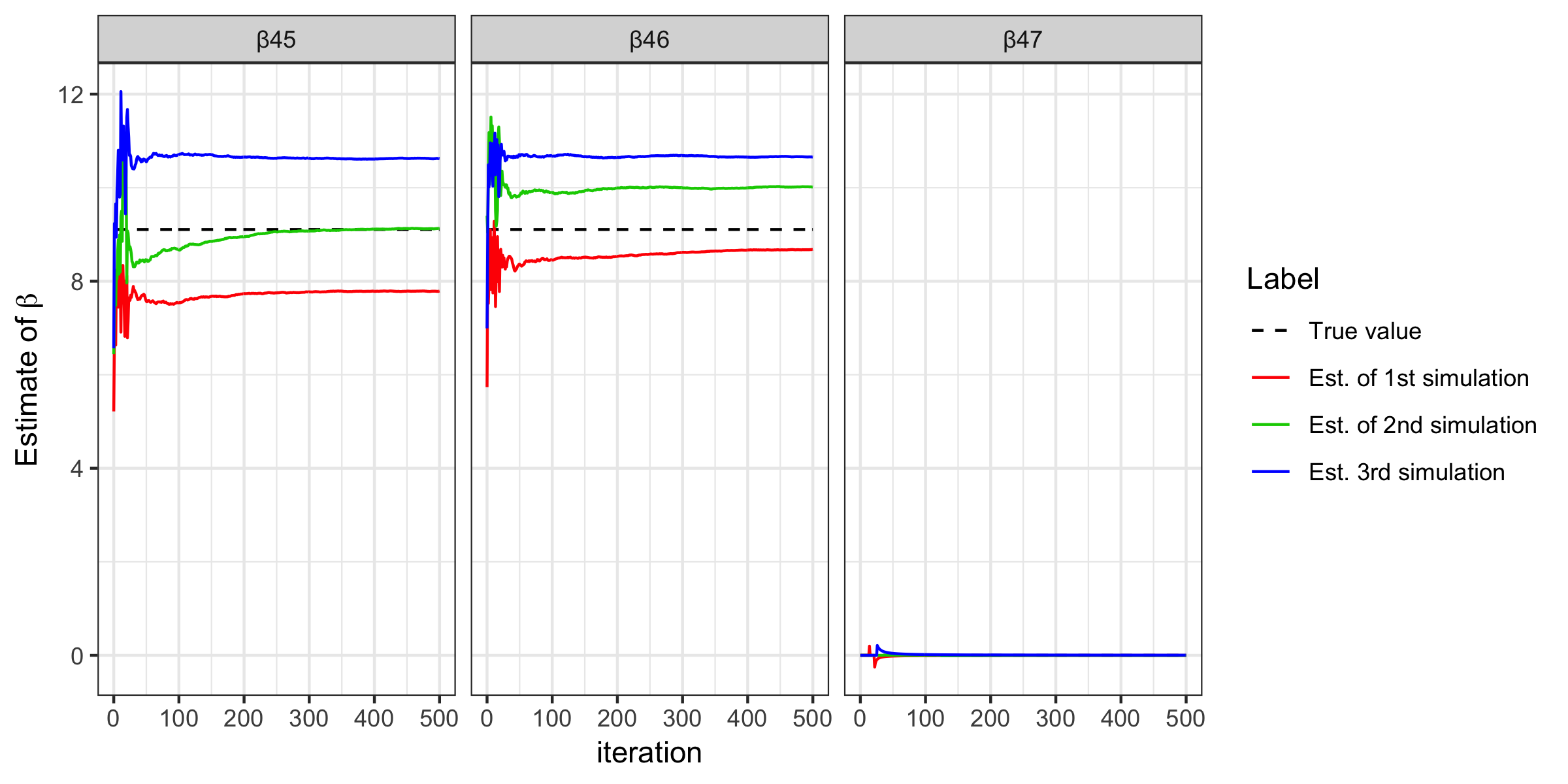}
\caption{Convergence plots for three coefficients with {ABSLOPE} (colored solid curves). Black dash lines represent the true value for each $\beta$.  Estimates obtained with three different sets of simulated data are represented by three different colors.}
\label{fig:converge}
\end{figure}

Figure \ref{fig:converge} shows the convergence of some coefficients with SAEM for three simulated data sets. These graphs are representative of all the observed results. There are large fluctuations during the first $t_0 = $20 iterations, then after introducing the stochastic approximation at the 20th iteration, convergence is achieved gradually. Due to the existence of a sorted  $l_1$ penalty, the estimates are still slightly biased. 

In addition, we also represent the convergence curves for $\sigma$ with {ABSLOPE} in supplementary materials \citep{sup} in order to compare the estimate of $\sigma$ by {ABSLOPE} to the biased MLE estimator without prior knowledge, \textit{i.e.}, $\hat{\sigma}_{\rm MLE}=\sqrt{\frac{RSS}{n}}$. We can see that the estimates of $\sigma$ with both methods are biased downward, but since {ABSLOPE} has an additional correction term (\ref{eq:sigma}), it leads to a less biased estimator.

\subsection{Behavior of ABSLOPE - SLOBE}\label{ssec:effect}
We then evaluate ABSLOPE and SLOBE in a different  parametrization setting  to see how the signal strength, sparsity and other parameters influence their performance.
\paragraph{Criterion}
We apply {ABSLOPE} or {SLOBE} on a synthetic dataset and get estimates for $\hat{\beta}$ and the sampled $\hat{\gamma}$ indicating the model selection results.
We compare the  selected model to the true one. 
The total number of true discoveries is $TP = \#\{j: \vert \beta_j \rvert >0 \text{ and } \vert  \hat{\beta}_j \rvert >0\}$ 
and  the total number of false discoveries is $FN = \#\{j: \vert \beta_j \rvert >0 \text{ and }  \hat{\beta}_j =0\}$.  

To evaluate the performance, we consider the following quantities:
\begin{itemize}
\item Power $= \frac{TP}{TP + FN}$;
\item FDR $= \frac{FP}{FP+TP} $ ;
\item MSE of $\beta$ (Relative $l_2$ norm error) $=\frac{\Vert \hat{\beta} - \beta \rVert^2}{\Vert \beta \rVert^2}$;
\item Relative prediction error $=\frac{\Vert X\hat{\beta} - X\beta \rVert^2}{\Vert X\beta \rVert^2}$.
\end{itemize}
For each set of parameters, we repeat the procedure 200 times: \textit{i)} data generation \textit{ii)} estimation and model selection with ABSLOPE/SLOBE \textit{iii)} evaluation with the criteria presented above and we compute the means over the 200 simulations. The simulations were implemented with parallel computing. 

\subsubsection{Scenario 1}\label{ssec:sce1}
We first consider $n=p=100$ and vary:
\begin{itemize}
\item sparsity: number of true signal $k= 5, \, 10, \, 15, \, 20$;
\item signal strength $\sqrt{2\log p}  \, ,2\sqrt{2\log p} \, ,3\sqrt{2\log p}  \, ,4\sqrt{2\log p} $;
\item percentage of missingness $  0.1, \,  0.2, \, 0.3$, generated randomly, i.e., MCAR;
\item correlation between covariates $\Sigma =
\text{toeplitz} (\rho)$\footnote{The Toeplitz structure (or auto-regressive structure) for correlation has been introduced for microarry study \citep{toeplitz_microarry}, with the form: $
\Sigma=\left( \begin{array}{ccccc}{1} & {\rho} & {\cdots} & {\rho^{p-2}} & {\rho^{p-1}} \\ {\rho} & {1} & {\ddots} & {\cdots} & {\rho^{p-2}} \\ {\vdots} & {\ddots} & {\ddots} & {\ddots} & {\vdots} \\ {\rho^{p-2}} & {\cdots} & {\ddots} & {\ddots} & {\rho} \\ {\rho^{p-1}} & {\rho^{p-2}} & {\cdots} & {\rho} & {1}\end{array}\right)_{p \times p}
$, where $\rho \in [0,1]$ is a constant. For the Toeplitz structure, adjacent pairs of covariates are highly correlated and those further away are less correlated, as in microarry study, genes are correlated due to their distance in the regularity pathway.} where $\rho=0, \,  0.5,\, 0.9$.
\end{itemize}
Then we applied the Algorithm \ref{alg:ABSLOPE} on each synthetic dataset. 
\paragraph{Results 1: no correlation, 10\% missingness - vary signal strength}
According to Figure \ref{fig:res1}:
\begin{figure}[!htbp]
\centering
\subfloat[Power]{\includegraphics[width=0.4\textwidth]{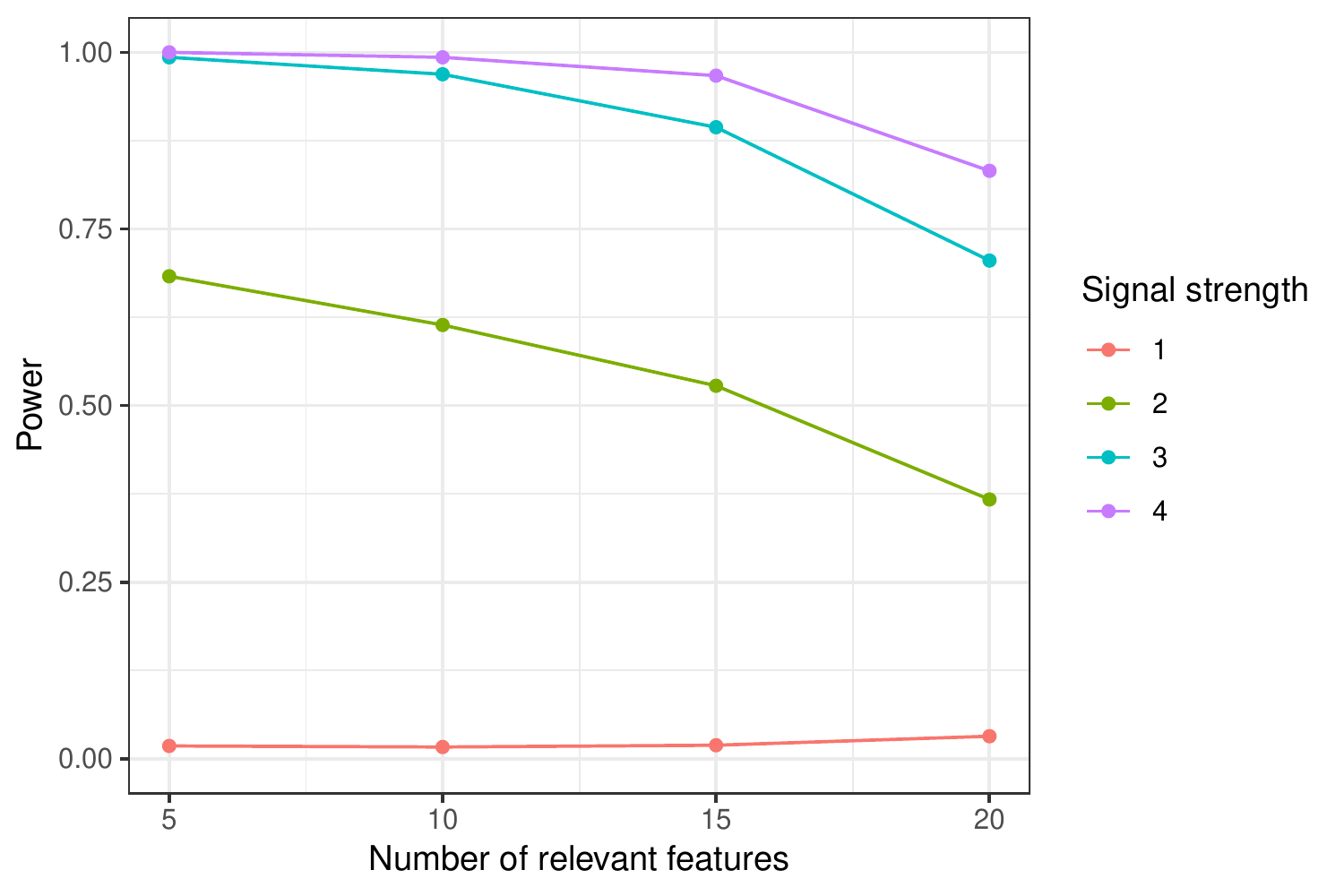}}
\subfloat[FDR]{\includegraphics[width=0.4\textwidth]{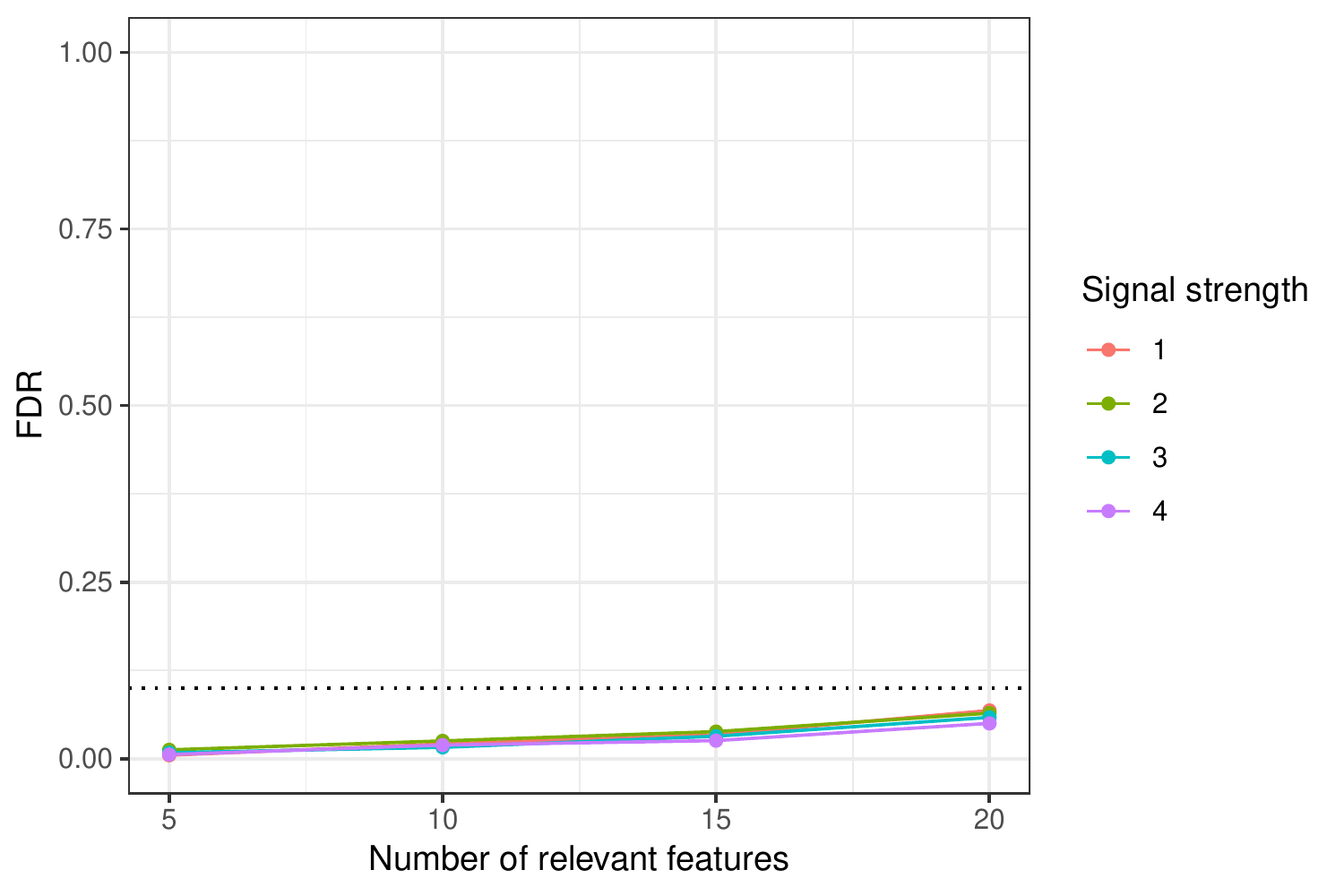}}\\
\subfloat[Bias of $\beta$]{\includegraphics[width=0.4\textwidth]{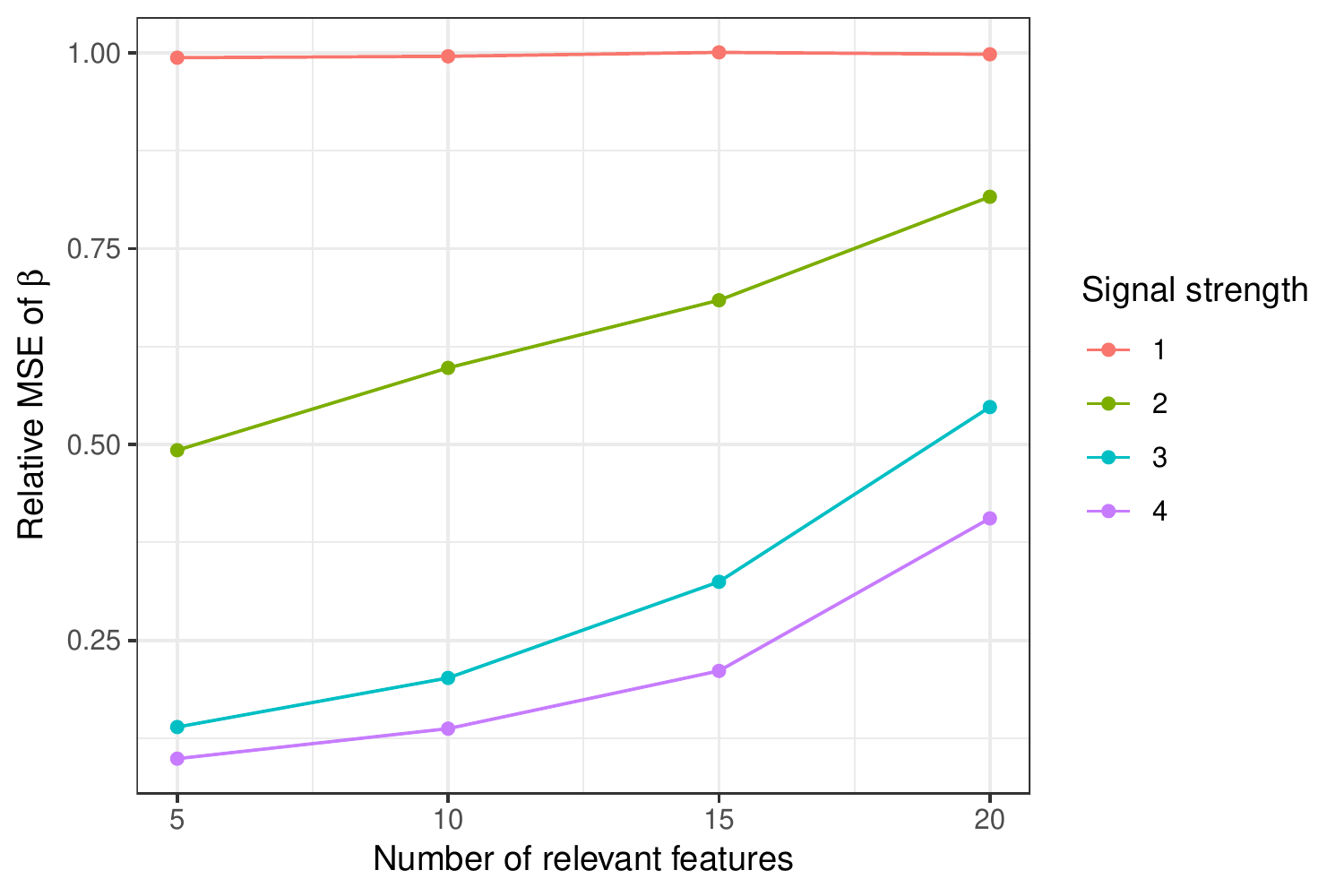}}
\subfloat[Prediction error]{\includegraphics[width=0.4\textwidth]{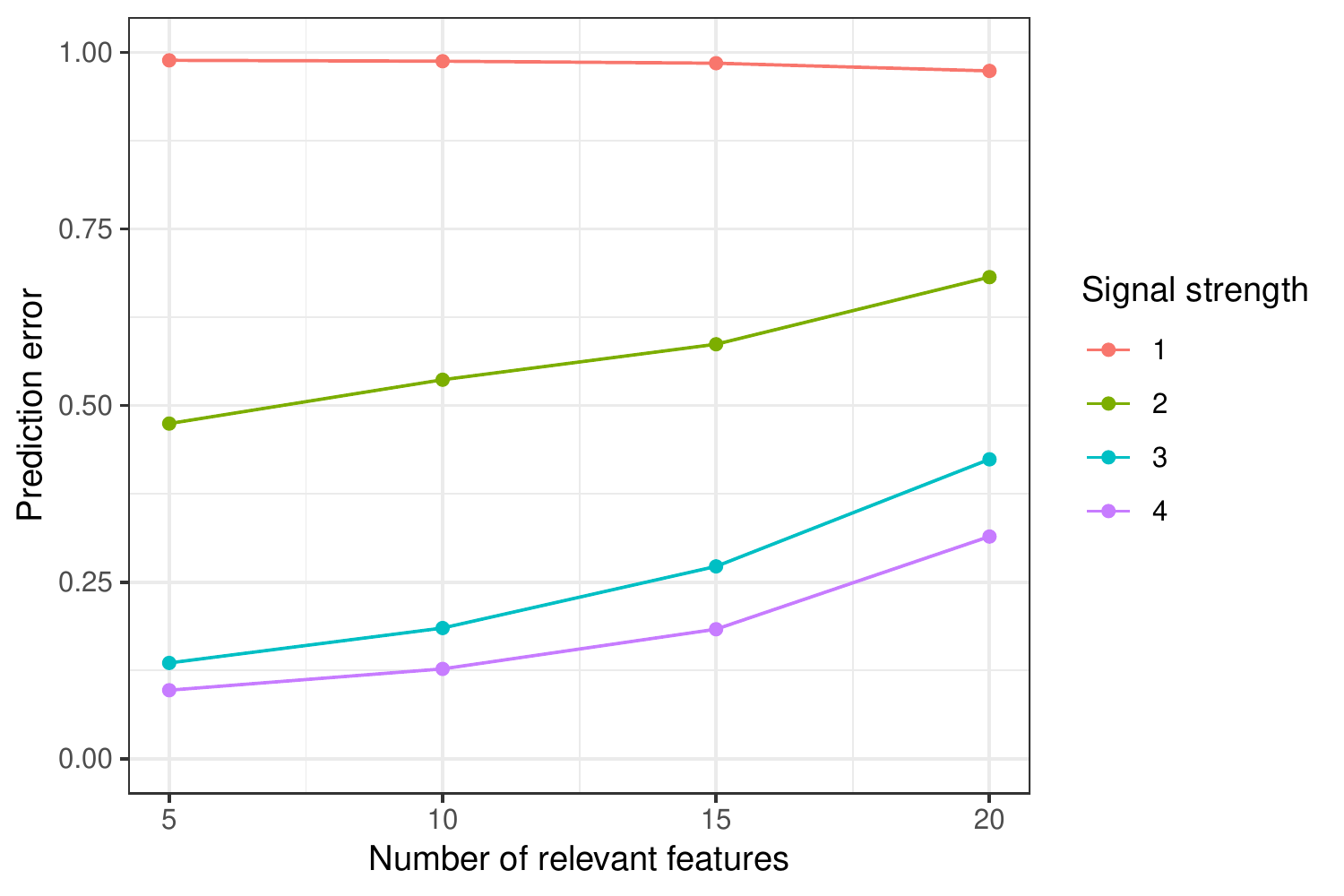}}
\caption{Mean of power (a), FDR (b), bias of the estimate for $\beta$ (c) and prediction error (d), as function of length of true signal, over the 200 simulations. Results for $n=p=100$, percentage of missingness $10\%$ and $\Sigma$ orthogonal (no correlation).}
\label{fig:res1}	
\end{figure}
\begin{itemize}
\item We observe that FDR is always controlled at the expected level 0.1.
\item Power increases and estimation bias decreases with larger sparsity or stronger signal.
\item When the signal is too weak (signal strength = $\sqrt{2\log p}$), the power is near 0, which is due to the identifiablility issue that ABSLOPE cannot distinguish the signal from the noise. Indeed, the value $c=\frac{\lambda_1}{\sigma \sqrt{2\log p}}$ is greater than one where $\lambda_1$ is the largest penalization coefficient. In addition, the bias is significant. This behaviour can be explained by the fact that we choose the penalty $\lambda$ to reduce the noise $\sigma$; but when the signal is as weak as $\sigma$, this choice of $\lambda$ also "kills" the real signal. 
\end{itemize}

\paragraph{Results 2: with correlation, strong signal - vary percentage of missingness}
Now we add the correlation as $\Sigma =
\text{toeplitz} (\rho)$ where $\rho=0.5$, and also fix a strong signal strength as $3\sqrt{2\log p}$. We then vary the sparsity and percentage of missingness. The results in Figure \ref{fig:res2} show that:
\begin{figure}[!htbp]
\centering
\subfloat[Power]{\includegraphics[width=0.4\textwidth]{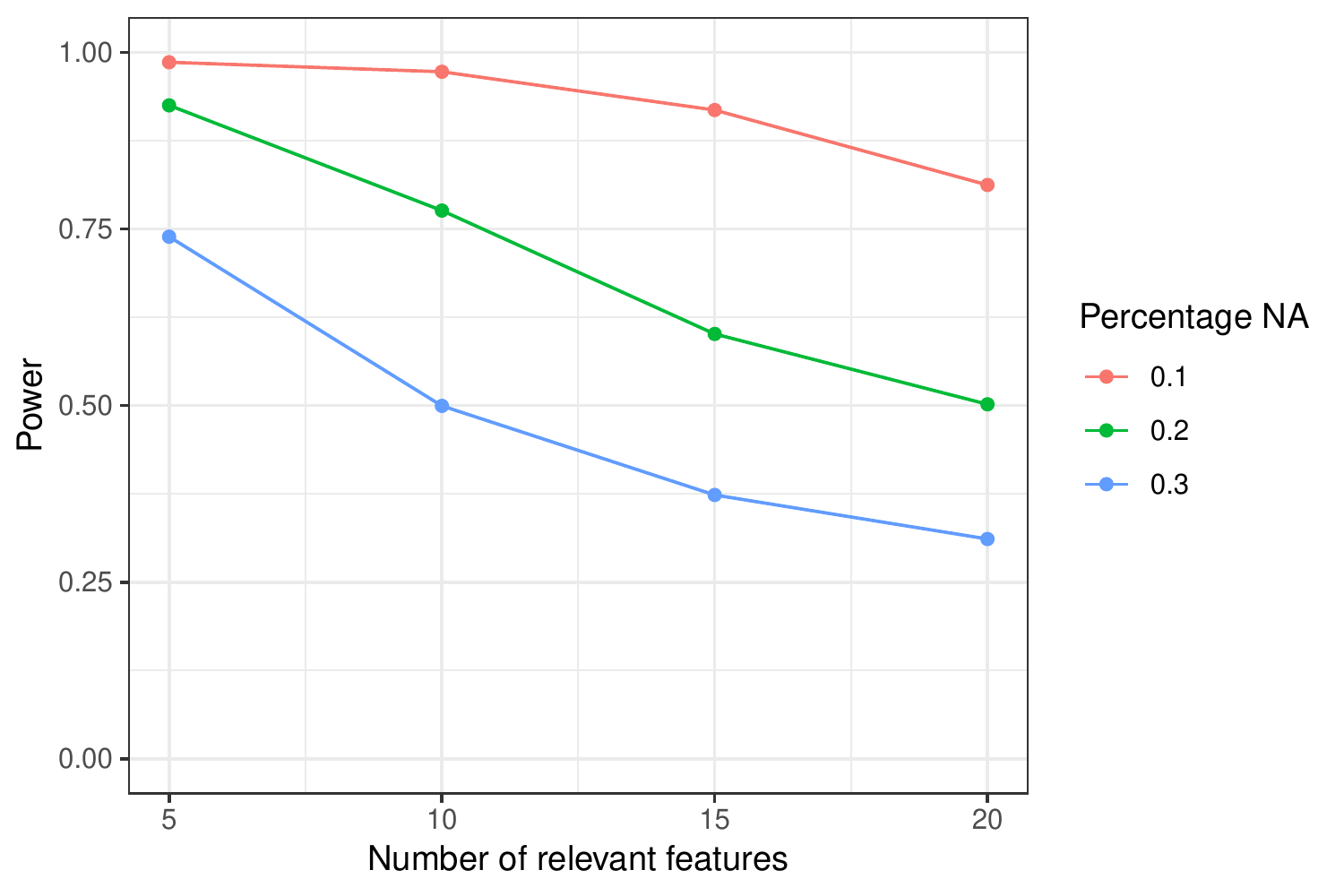}}
\subfloat[FDR]{\includegraphics[width=0.4\textwidth]{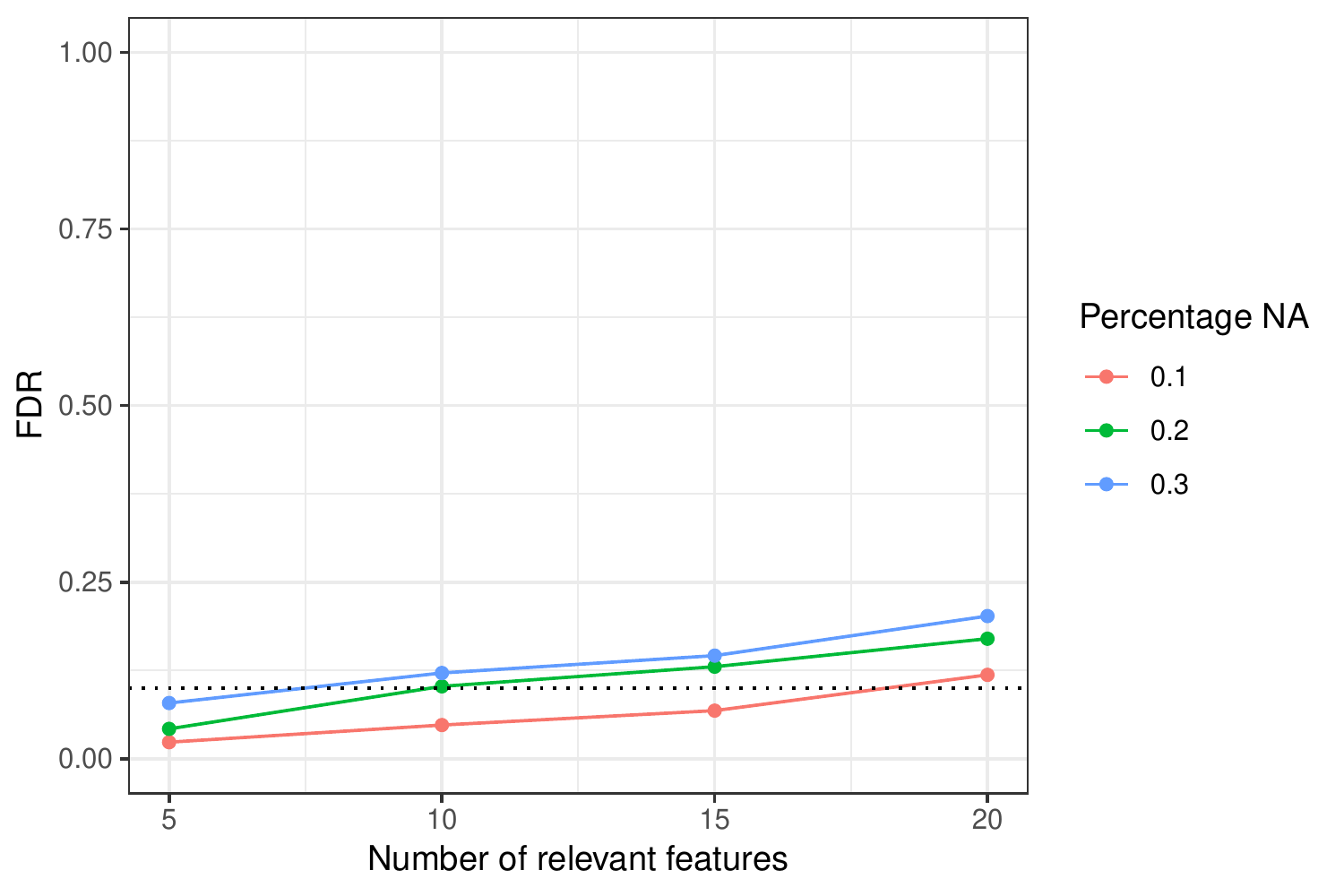}}\\
\subfloat[Bias of $\beta$]{\includegraphics[width=0.4\textwidth]{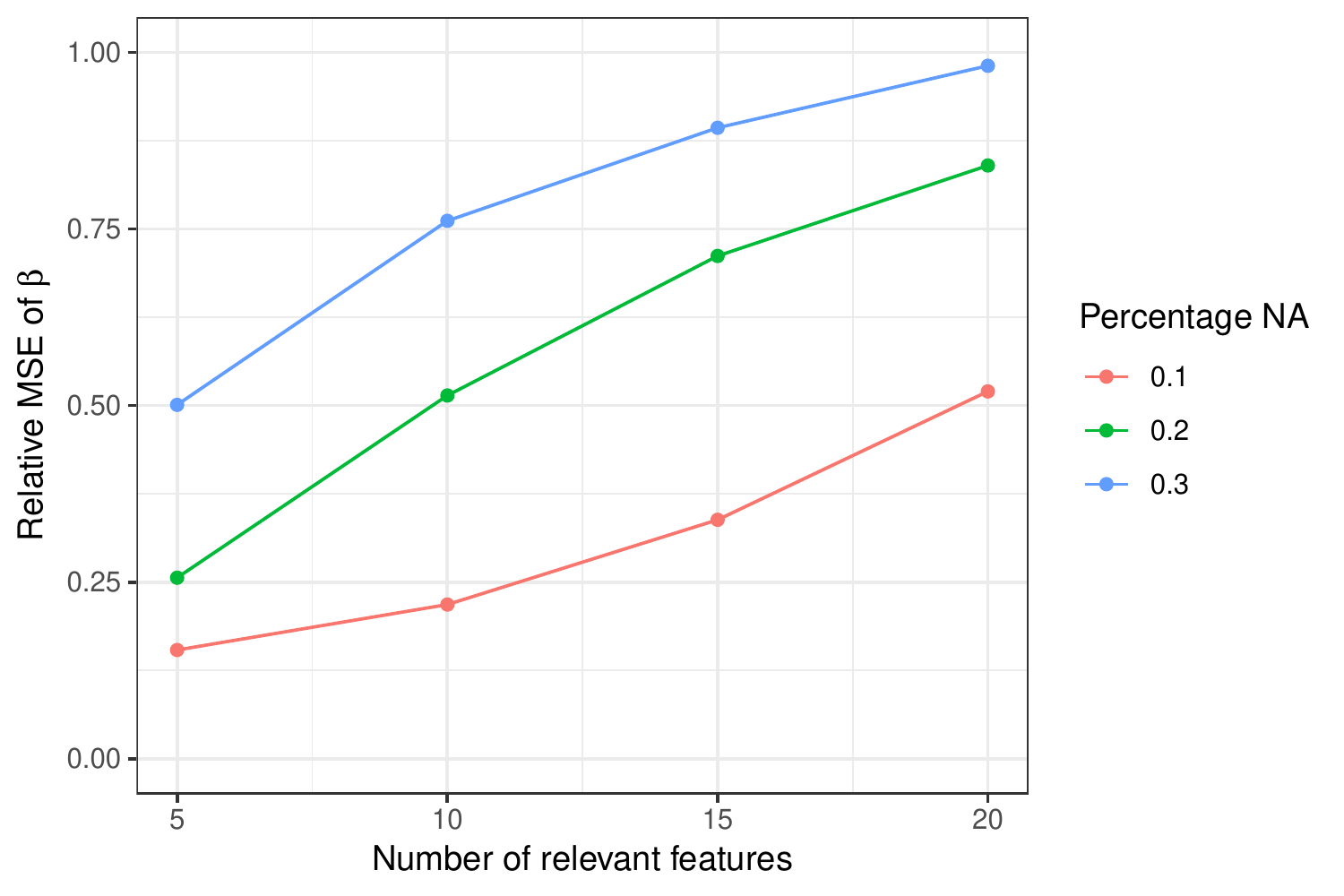}}
\subfloat[Prediction error]{\includegraphics[width=0.4\textwidth]{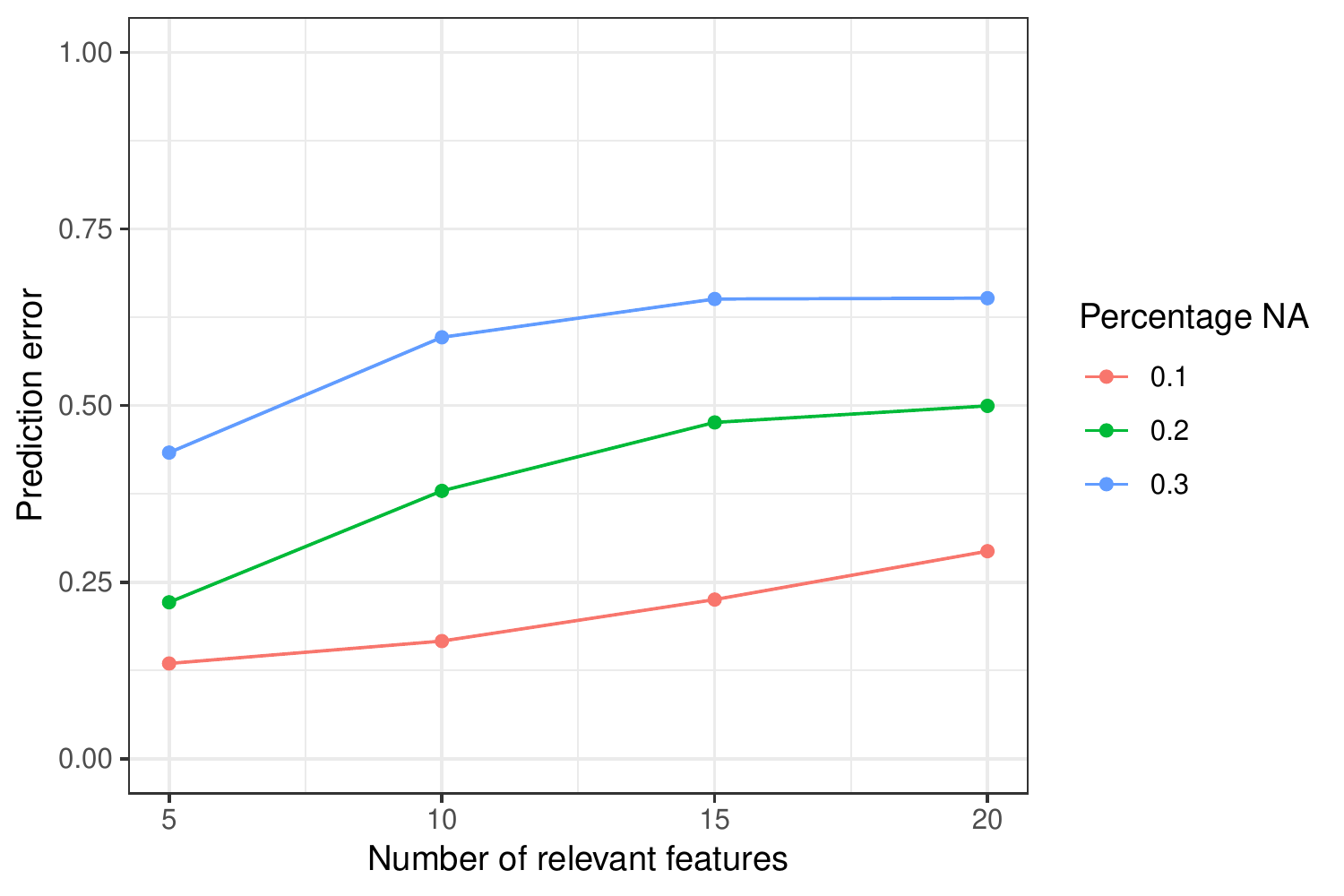}}
\caption{Mean of power (a), FDR (b), bias of the estimate for $\beta$ (c) and prediction error (d), as function of length of true signal  over the 200 simulations. Results for $n=p=100$, with correlation and strong signal.}
\label{fig:res2}	
\end{figure}
\begin{itemize}
\item The power increases and the estimation bias decreases when the percentage of missing data decreases.
\item In the presence of correlation, the FDR control is slightly lost when the number of non-zero coefficients is greater than 10 and the percentage of missing values exceeds 0.2, but is still  near the nominal level.
\end{itemize}

\subsubsection{Scenario 2}\label{ssec:sce2}
Now we consider a larger dataset $n=p=500$ and vary the same parametrization as in Subsection \ref{ssec:sce1}, except the sparsity, for which we take wider range of choices among $k= 10, \, 20, \, 30, \, \cdots, \, 60$. In this scenario of larger dimension, we have applied the simplified {SLOBE} algorithm as described in Subsection \ref{ssec:slob} to avoid intensive computation.
\begin{figure}[!htbp]
\centering
\subfloat[Power]{\includegraphics[width=0.4\textwidth]{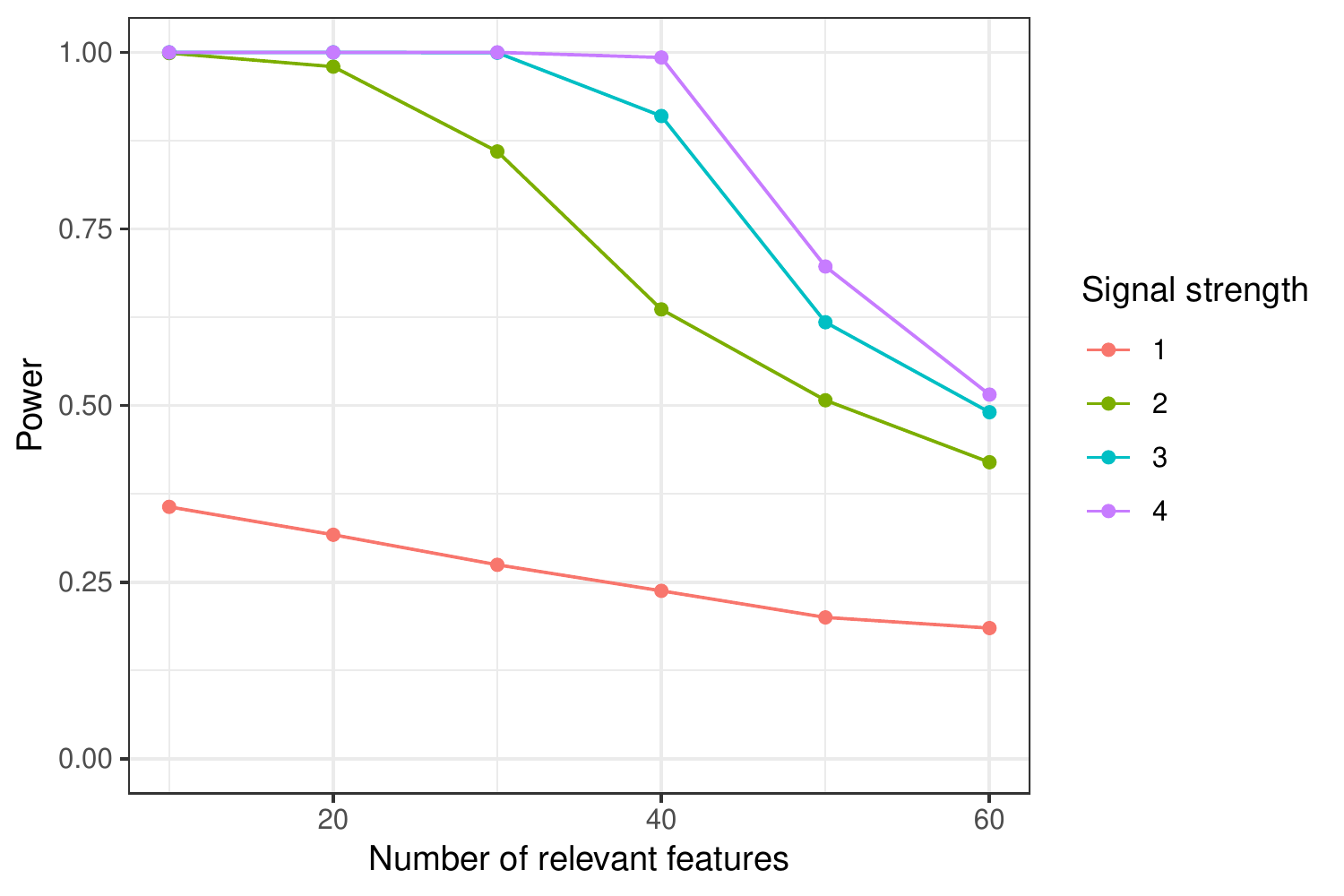}}
\subfloat[FDR]{\includegraphics[width=0.4\textwidth]{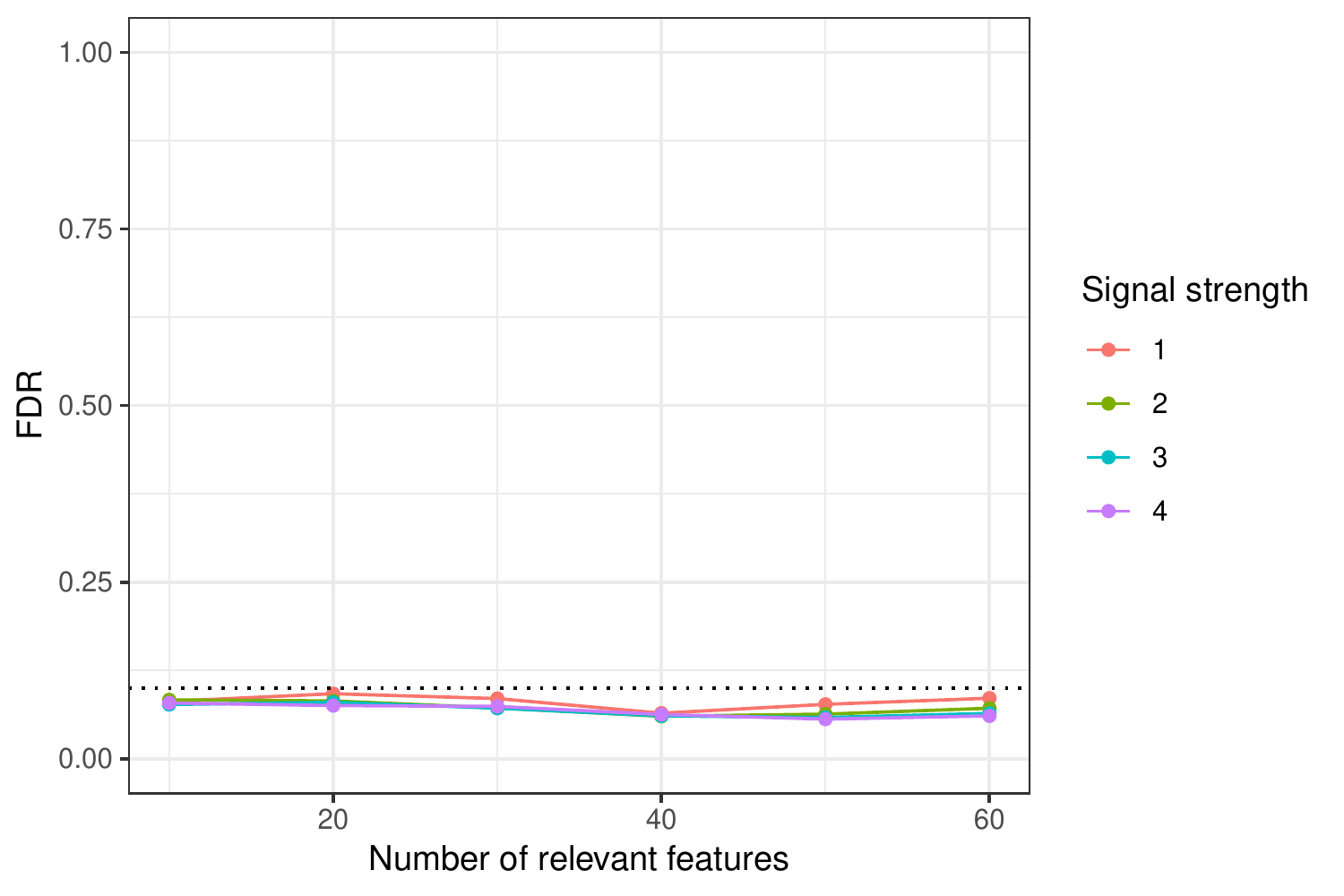}}\\
\subfloat[Bias of $\beta$]{\includegraphics[width=0.4\textwidth]{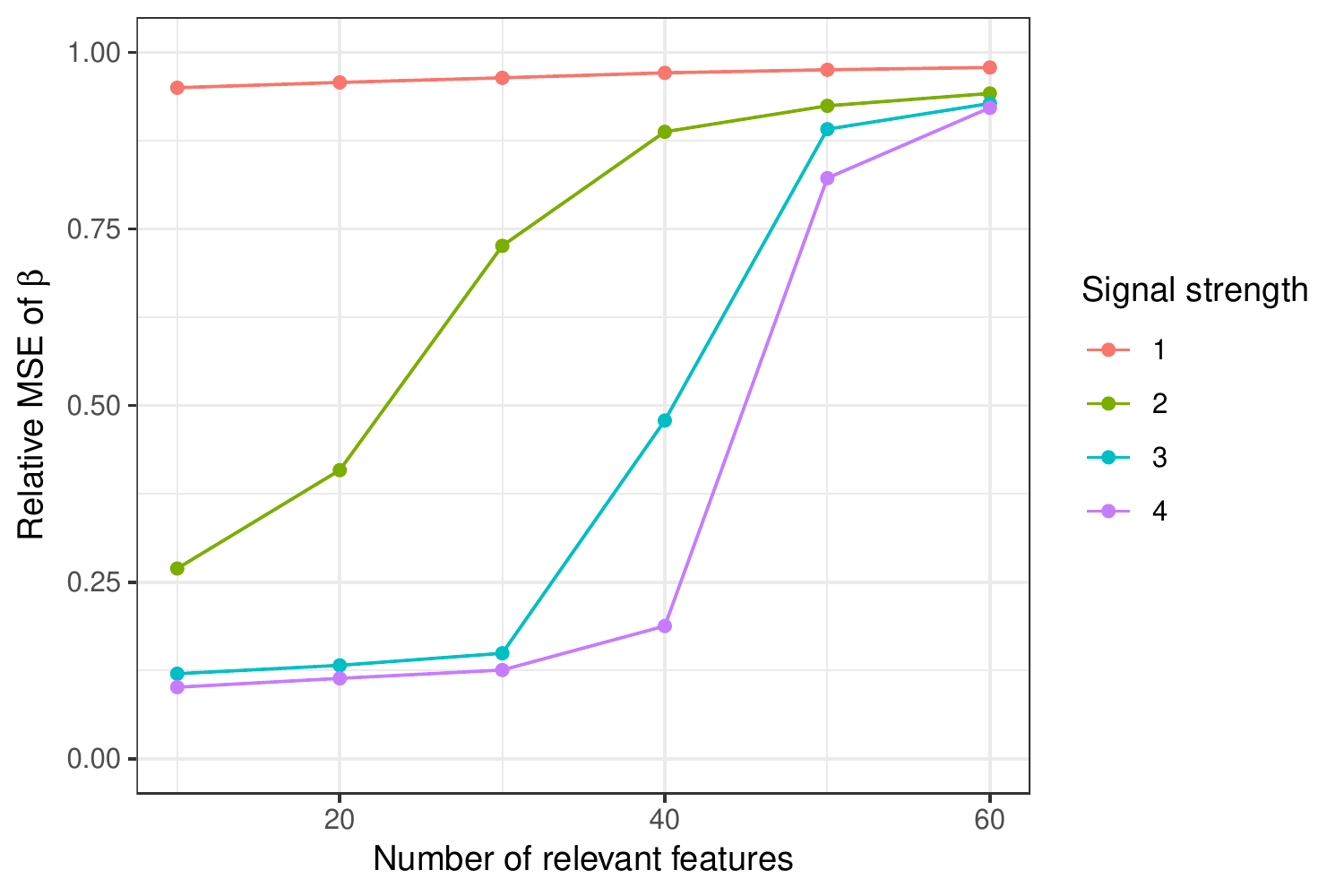}}
\subfloat[Prediction error]{\includegraphics[width=0.4\textwidth]{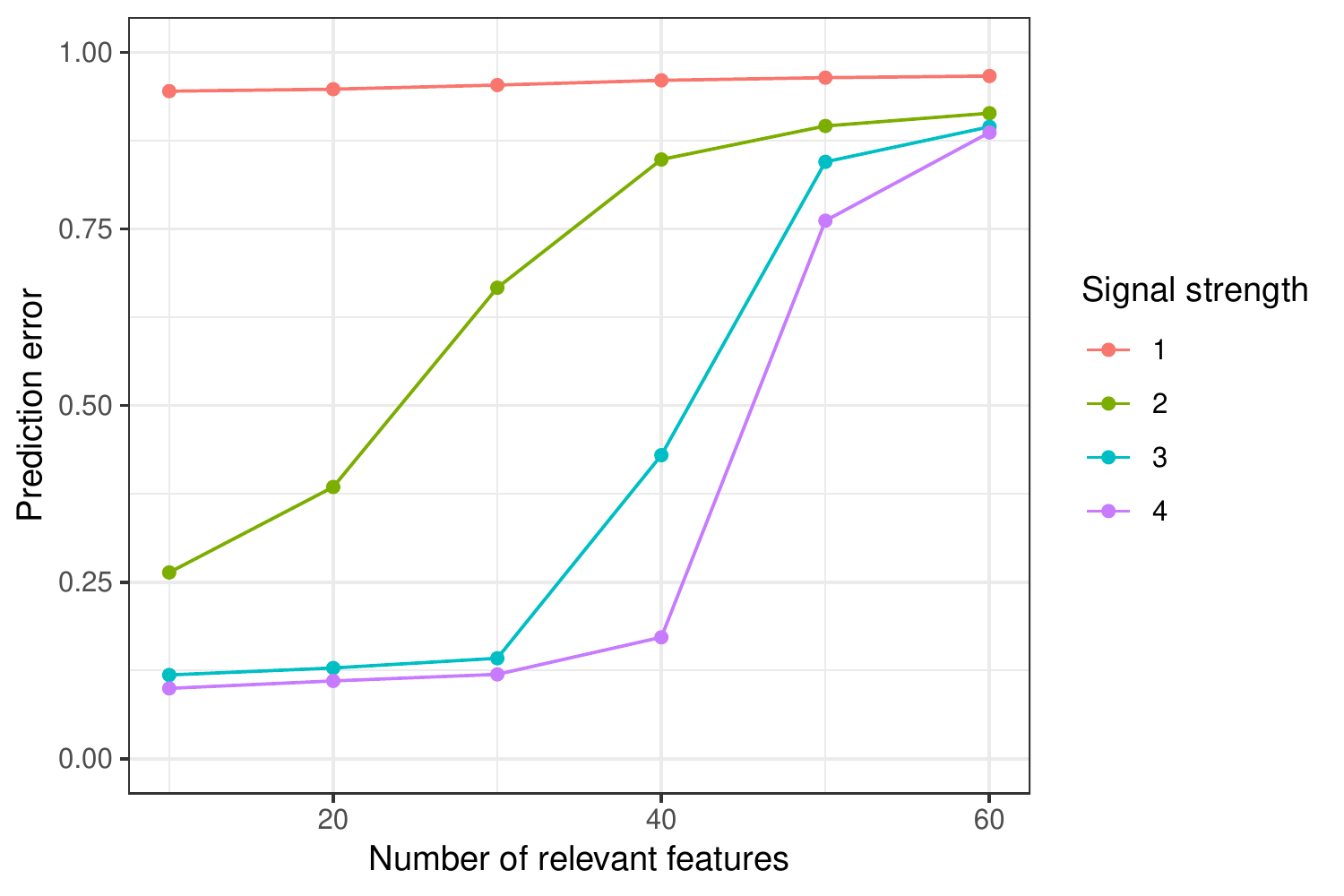}}
\caption{Mean of power (a), FDR (b), bias of the estimate for $\beta$ (c) and prediction error (d), as function of length of true signal, over the 200 simulations. Results for $n=p=500$, percentage of missingness $10\%$ and $\Sigma$ orthogonal (no correlation).}
\label{fig:500res1}	
\end{figure}
 \paragraph{Results 1: no correlation, 10\% missingness - vary signal strength}
According to Figure \ref{fig:500res1}:

\begin{itemize}
\item FDR is always controlled at expected level 0.1.
\item Similar to Figure \ref{fig:res1}, power increases and estimation error decreases with larger sparsity and stronger signal. However in this larger dimension case, we can handle with larger number of relevant features until 30 or 40, at which we observe a phase transition due to the identifiability issue. 
\end{itemize}

\paragraph{Results 2: with correlation, strong signal - vary percentage of missingness}
Now we add the correlation as $\Sigma =
\text{toeplitz} (\rho)$ where $\rho=0.5$, and also fix a strong signal strength as $3\sqrt{2\log p}$. We then vary the sparsity and percentage of missingness. The results in Figure \ref{fig:500res2} show that:
\begin{figure}[!htbp]
\centering
\subfloat[Power]{\includegraphics[width=0.38\textwidth]{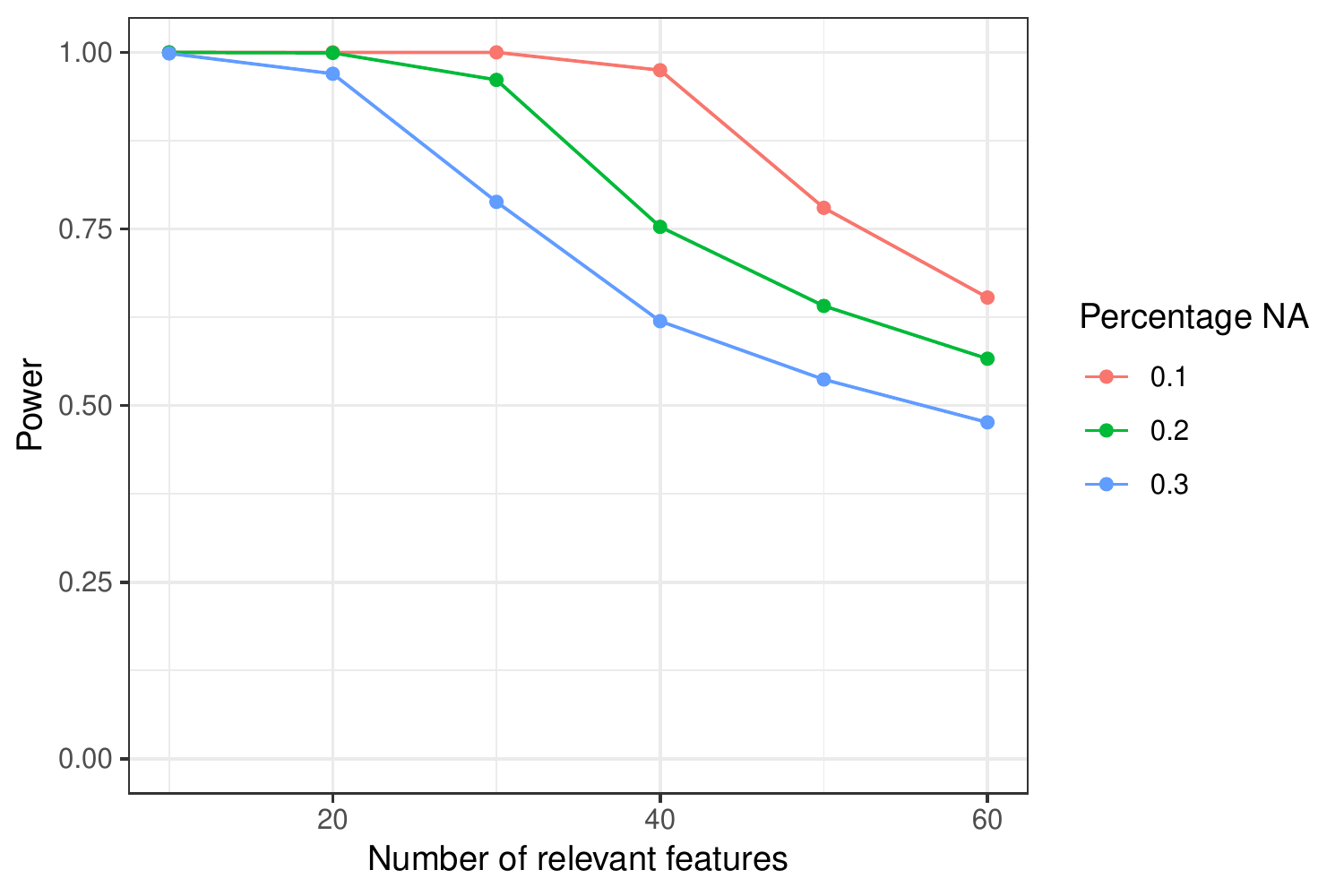}}
\subfloat[FDR]{\includegraphics[width=0.38\textwidth]{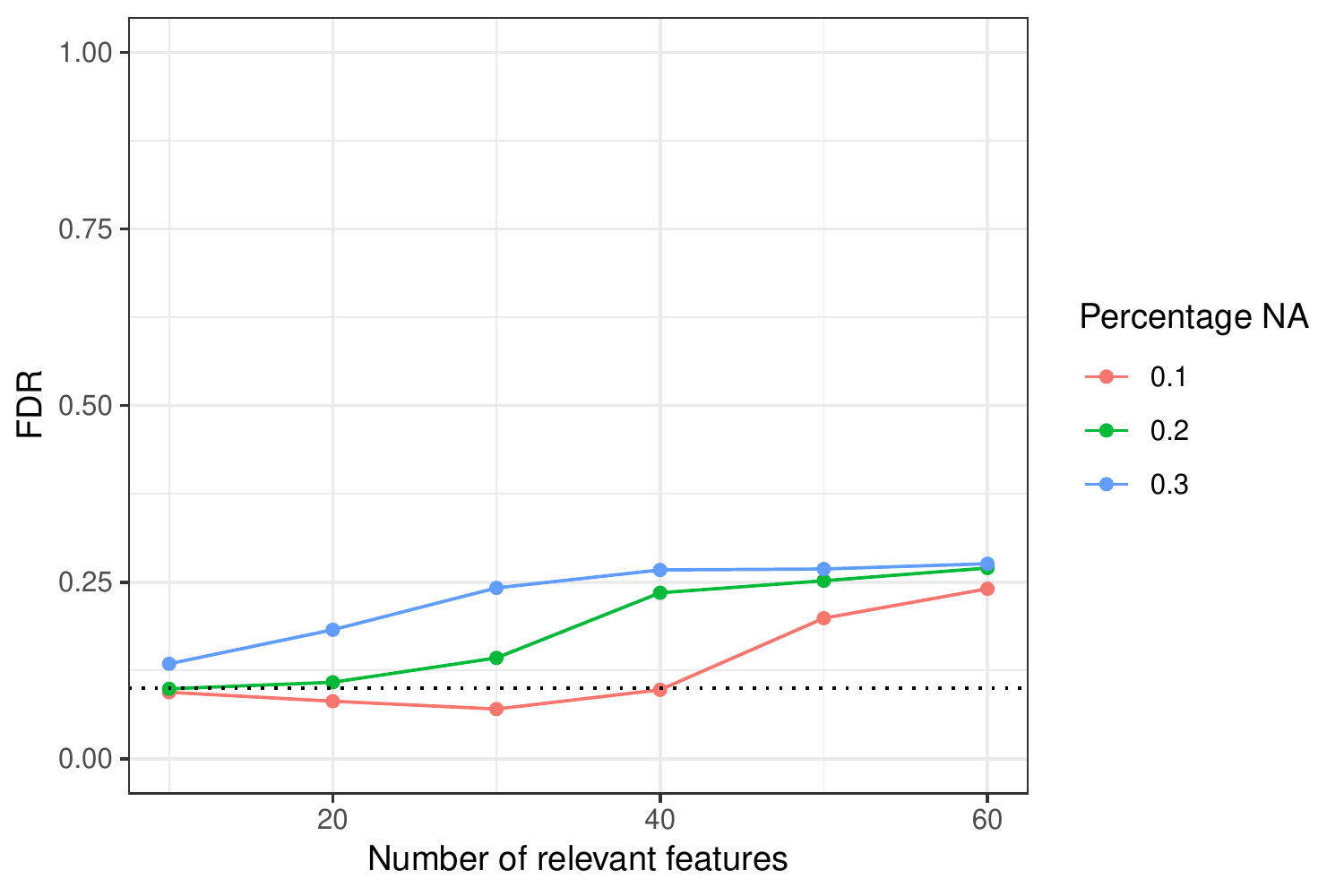}}\\
\subfloat[Bias of $\beta$]{\includegraphics[width=0.38\textwidth]{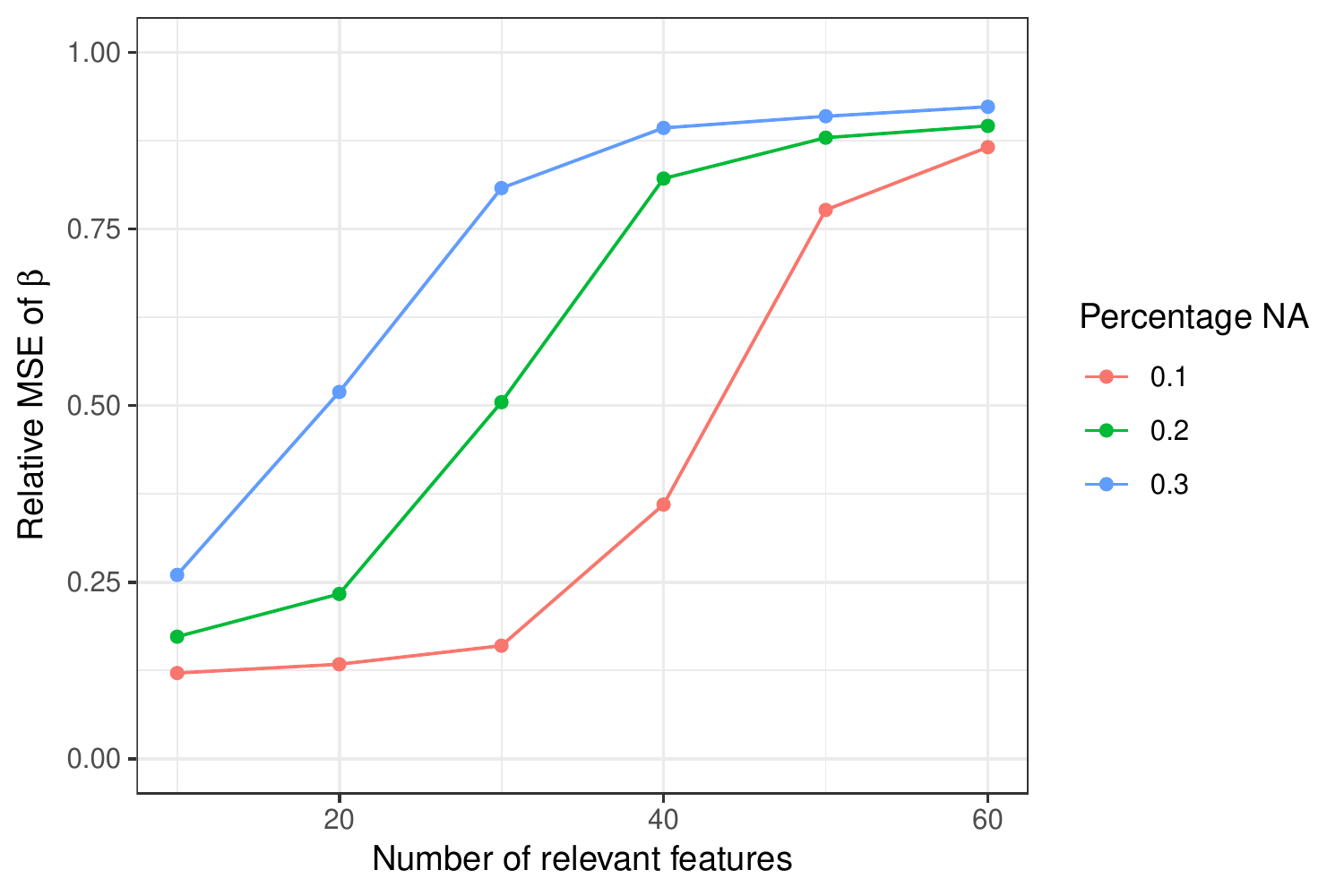}}
\subfloat[Prediction error]{\includegraphics[width=0.38\textwidth]{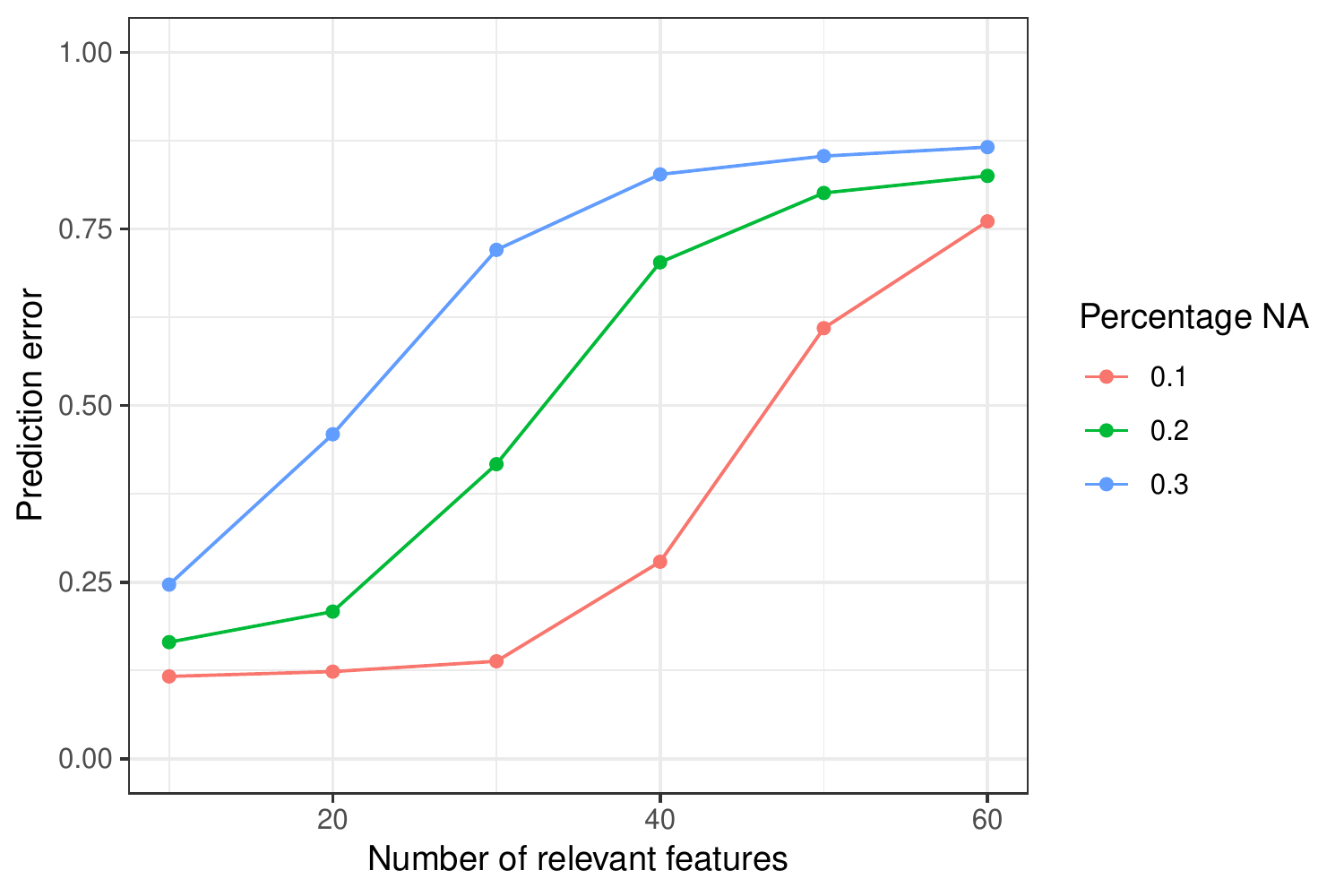}}
\caption{Mean of power (a), FDR (b), bias of the estimate for $\beta$ (c) and prediction error (d), as function of length of true signal  over the 200 simulations. Results for $n=p=500$, with correlation and strong signal.}
\label{fig:500res2}	
\end{figure}
\begin{itemize}
\item Similar to Figure \ref{fig:res2}, the power increases and the estimation error decreases when the percentage of missing data decreases.
\item Due to the existence of correlation, the FDR control is slight lost, especially in the less sparse and more missing case.
\item With 10\% missing values, if the number of relevant features is below 40, then we can always achieve an efficient power and perfect FDR control. With larger percentage of missing values, the sparsity of this changing point will be more conservative.
\end{itemize}

In addition, we present the results varying the correlations in the supplementary materials \citep{sup}.

\subsection{Comparison with competitors}
 We use the same simulation scenario and criteria as those used in Subsection \ref{ssec:effect} to compare {ABSLOPE} and {SLOBE} to other approaches that can be considered to select variables in the presence of missing data.
 
\begin{itemize}
\item ncLASSO: Non-convex LASSO \citep{loh2012}
\item Methods based on preliminary mean imputation (MeanImp): missing values are replaced by the average of the observed values for each variable, then on the completed data set is applied:
\begin{itemize}
\item SLOPE: Applying two steps \textit{i)} SLOPE \citep{bogdan2015} \textit{ii)} OLS  on the selected predictors to estimate the parameters;
\item LASSO: LASSO with $\lambda$ selected by cross validation; 
\item adaLASSO: adaptive LASSO \citep{zou2006adaptive};
\end{itemize}
\end{itemize}
For SLOPE, {ABSLOPE} and {SLOBE}, we set the penalization coefficient $\lambda$ as the BH sequence which controls the FDR at level $0.1$.
The values of the tuning parameters for the different methods can be found in the available code on GitHub \citep{github}. We try to make the comparisons as fair as possible and also favor the competitors: we give the true $\sigma$ to SLOPE whereas we estimate it with { ABSLOPE}. ncLASSO requires to specify  a bound on the $l_1$ norm of the coefficients, \textit{i.e.}, $\beta<R = b_0 \#\{\beta_j: \beta_j \neq 0\}$, for which we take the real value of sparsity and signal strength.

Note that we do not make comparisons with the widely used multiple imputation \citep{mice}, where several imputed values are made for each missing value to reflect the uncertainty in the missingness. There are several reasons, including the inability to perform model selection with multiple imputation and the difficulty to aggregate the estimates from the imputed datasets.

We present the results for the case $n=p=100$ in the supplementary materials \citep{sup} while Figure \ref{fig:n500p500}  summarizes the result for the case  $n=p=500$, 10\% missingness and with correlation $\text{toeplitz}(0.5)$. Lighter colors indicate smaller values.
\begin{figure}[!htbp]
\centering
\subfloat[Power]{\includegraphics[width=0.4\textwidth]{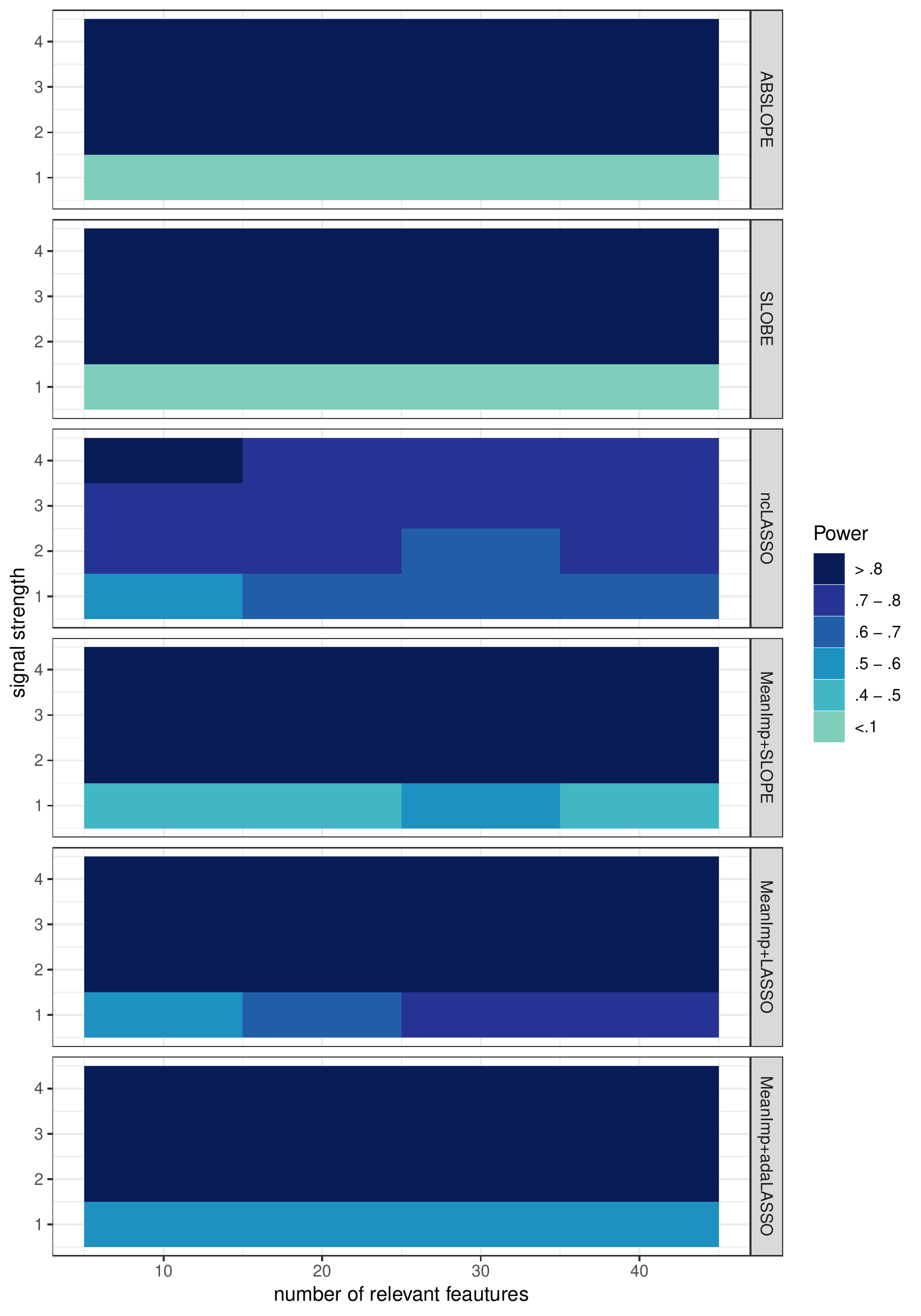}}
\hfill
\subfloat[FDR]{\includegraphics[width=0.4\textwidth]{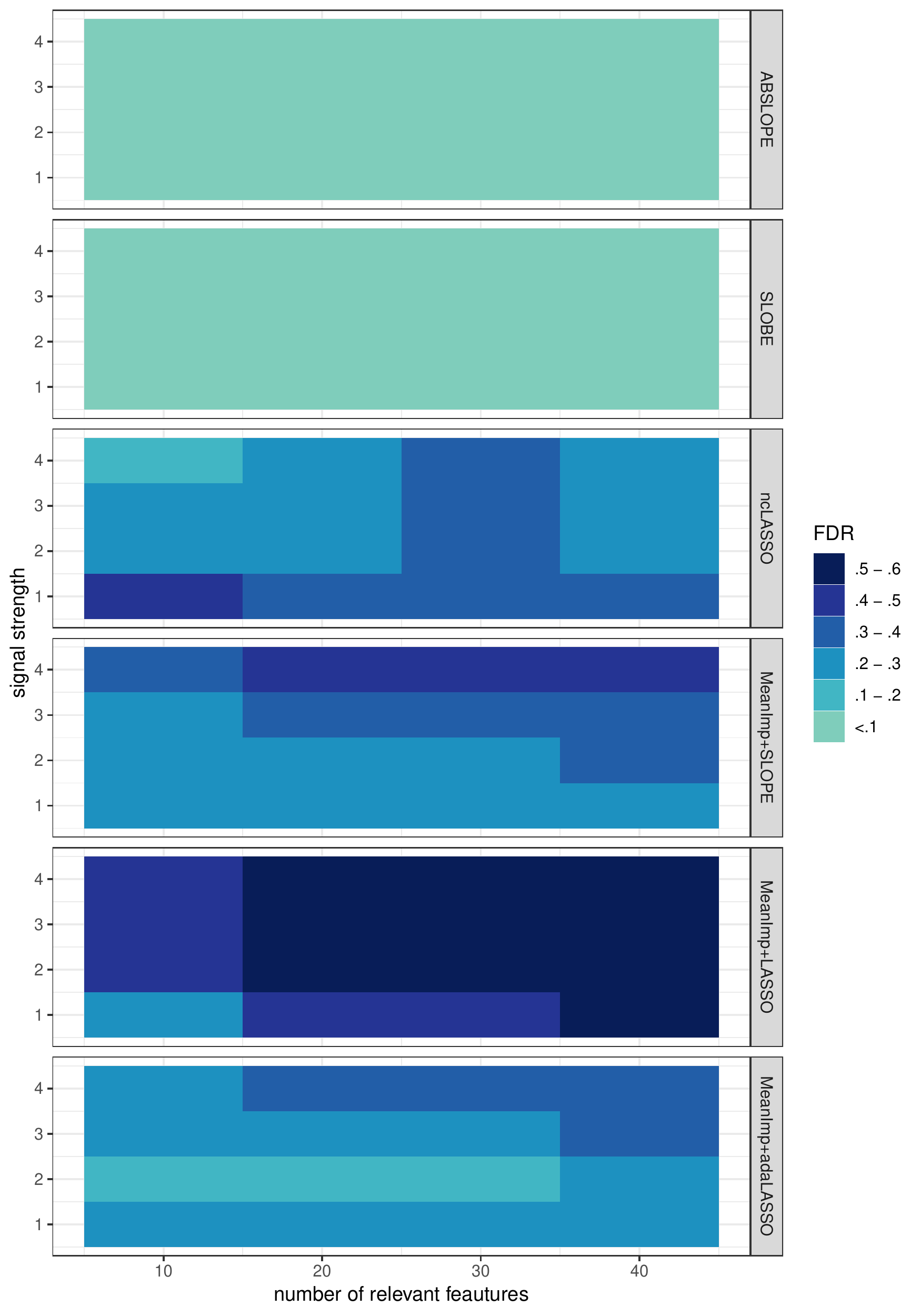}}\\
\subfloat[Bias of $\beta$]{\includegraphics[width=0.4\textwidth]{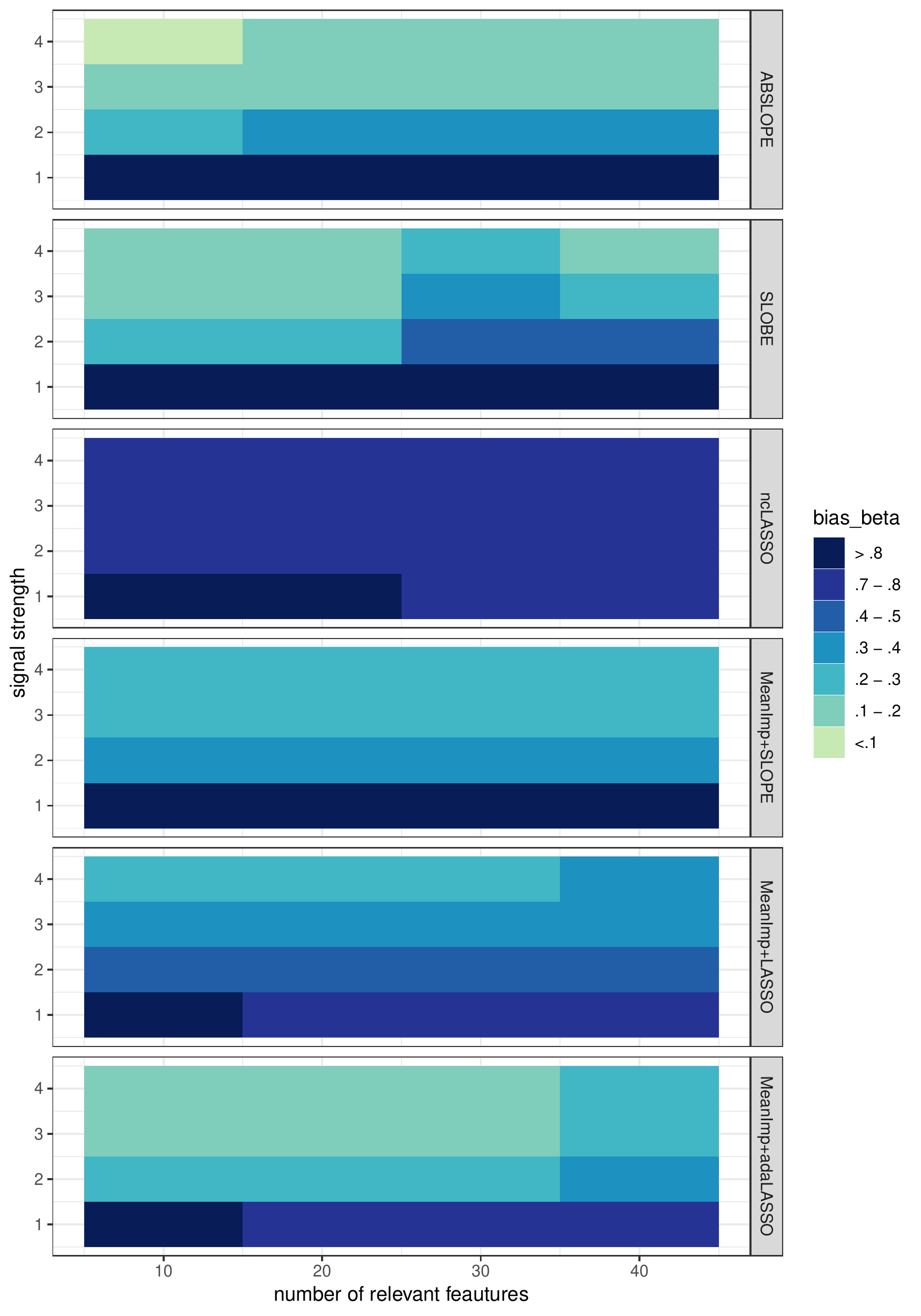}}
\hfill
\subfloat[Prediction error]{\includegraphics[width=0.4\textwidth]{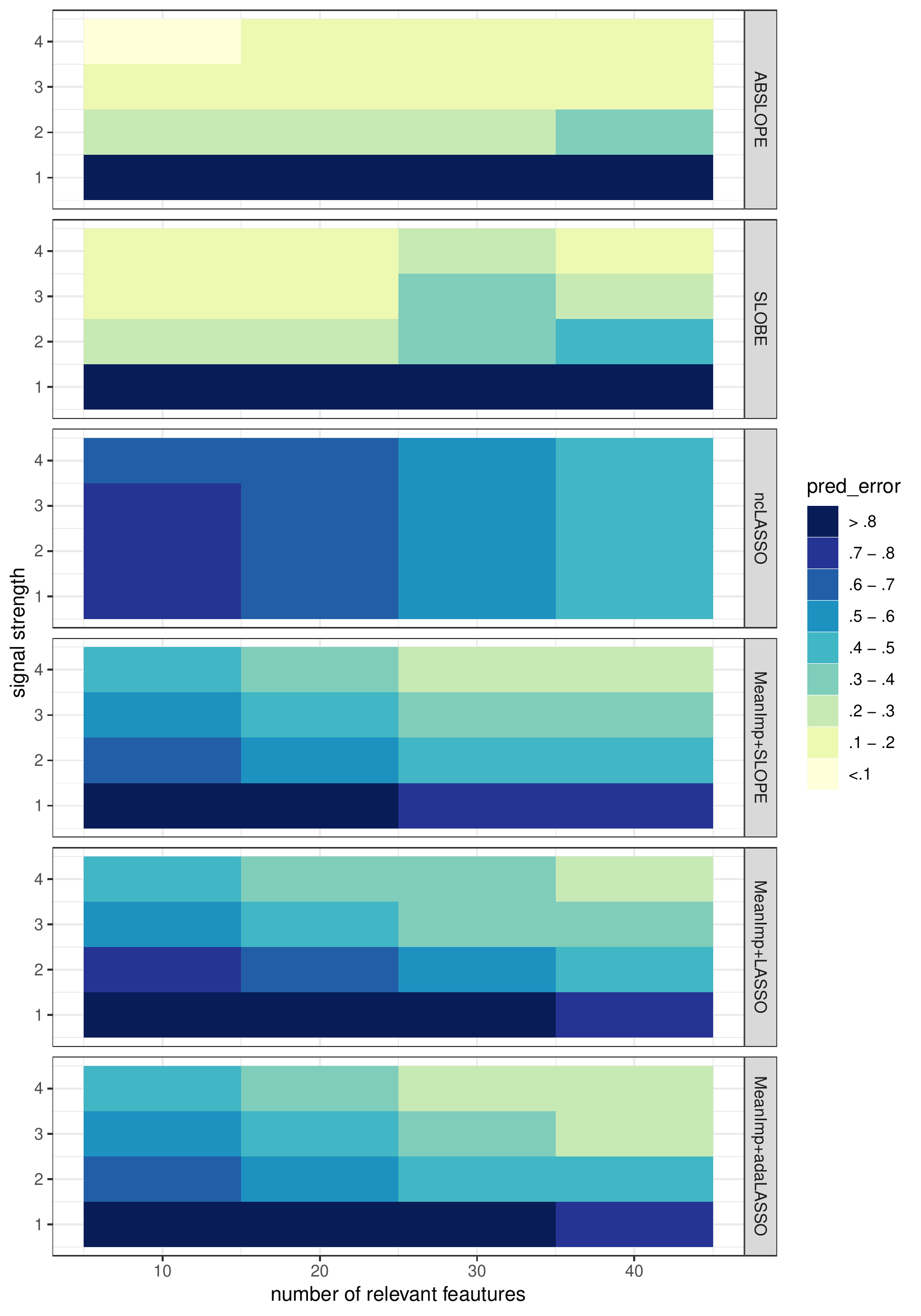}}
\caption{Comparison of power (a), FDR (b), bias of $\beta$ (c) and prediction error (d) with varying sparsity and signal strength, with 10\% missingness over 200 simulations in the case with correlation.}
\label{fig:n500p500}
\end{figure} 
\begin{itemize}
\item {ABSLOPE} and {SLOBE} both have strong power and  accurate prediction, where FDR is always controlled.
\item The power and FDR control achieved by {ABSLOPE} and {SLOBE} are better than the case $n=p=100$. On one hand, correlation helps the generation of missing values.
On the other hand, sparsity considered here is less complicated.
\item Other methods pay the price of FDR control to achieve good power. 
\end{itemize}

\subsection{Comparison of computation time}\label{ssec:time}
 Table \ref{tb:time} presents the execution time of the different methods considered in the simulation. In addition, we have  implemented  our proposed algorithm in C and we use Rcpp \citep{rcpp} to integrate these functions within R.
\begin{table}[!htbp]
\caption{\label{tb:time} Comparison of average execution time (in seconds) for one simulation, in the case without correlation and with $10\%$ MCAR, for $n=p=100$ and $n=p=500$ calculated over 200 simulations. (MacBook Pro, 2.5 GHz, processor Intel Core i7) }
\centering
\fbox{
\begin{tabular}{l|ccc|ccr}
  Execution time (seconds) & \multicolumn{3}{c|}{$\boldsymbol{n=p=100}$} & \multicolumn{3}{c}{$\boldsymbol{n=p=500}$} \\ for one simulation & min  & mean  & max & min  & mean  & max \\ 
  \hline
ABSLOPE  & 12.83 &  14.33 &  20.98    & 646.53 & 696.09 & 975.73 \\
SLOBE & 0.53  & 0.60   & 0.98  & 35.82 & 39.18 & 57.66 \\
\textbf{SLOBE (with Rcpp)} & 0.31 & 0.34 & 0.66 & 14.23 & 15.07 & 29.52 \\ 
MeanImp + SLOPE  & 0.01 & 0.02 &  0.09 & 0.24 & 0.28 & 0.53 \\
ncLASSO & 16.38 & 20.89 & 51.35 & 91.90 & 100.71 & 171.00\\
MeanImp + LASSO & 0.10 & 0.14 & 0.32 & 1.75 & 1.85 & 3.06 \\
MeanImp + adaLASSO  & 0.45 & 0.58 & 1.12  & 45.06 & 47.20 & 71.24\\
\end{tabular}}
\end{table}
In the case $n=p=100$, we observe that the most time consuming method is ncLASSO, which spent on average 20 seconds on one simulation. While ABSLOPE also took on average 14 seconds for one run, its simplified version SLOBE reduced this cost to 0.6 seconds, which is comparable to MeanImp + adaLASSO. While when $n=p=500$, the convergence of ABSLOPE requires much more time but SLOBE helps to simplify the complexity.
In addition, the version of C for SLOBE is more accelerated, saving half of the computation time, which makes SLOBE capable of handling  larger datasets.

\section{Application to Traumabase dataset}\label{sec:real}
\subsection{Details on the dataset and preprocessing}
Our work is motivated by an ongoing collaboration 
with the TraumaBase group\footnote{\url{http://www.traumabase.eu/}} at APHP (Public Assistance - Hospitals of Paris), which is dedicated to the management of severely traumatized patients. Major trauma is defined as any injury that endangers  life or  functional integrity of a person. The WHO has recently shown that major trauma in its various forms, including traffic accidents, interpersonal violence, self-harm, and falls, remains a public health challenge and a major source of mortality and handicap around the world \citep{20171260}. Effective and timely management of trauma is critical to improving outcomes. Delays and/or errors in treatment have a direct impact on survival, especially for the two main causes of death in major trauma: hemorrhage and traumatic brain injury. 
Using patients' records measured in the prehospital stage or on arrival to the hospital, we aim to establish prediction models in order to prepare an appropriate response upon arrival at the trauma center, \textit{e.g.}, massive transfusion protocol and/or immediate haemostatic procedures. 
Such models intend to give support to clinicians and professionals. 
Due to the highly stressful  multi-player environment, evidence suggests that patient management -- even in mature trauma
systems -- often exceeds acceptable time frames \citep{hamada2014evaluation}. In addition, discrepancies may be observed between diagnoses made by emergency doctors in the ambulance and those made when the patient arrives at the trauma center \citep{Hamada2015EuropeanTG}. These discrepancies can  result in poor outcomes such as inadequate hemorrhage control or delayed transfusion.

To improve decision-making and patient care, six trauma centers within the Ile de France region (Paris area) in France have collaborated to collect detailed high-quality clinical data from accident scenes to the hospital. These centers have joined TraumaBase progressively between January 2011 and June 2015. The database integrates algorithms for consistency and coherence and  data monitoring is performed by a central administrator. Sociodemographic, clinical, biological and therapeutic data (from the prehospital phase to the discharge) are systematically recorded for all trauma patients, and all patients transported in the trauma rooms of the participating centers are included in the registry. The resulting database now has data from 7495 trauma cases with more than 250 variables, collected from January 2011 to March 2016, with age ranged from 12 to 96, and is continually updated. The granularity of collected data makes
this dataset unique in Europe.  However, the data is highly heterogeneous, as it comes from multiple sources and, furthermore, is plagued with missing values, which makes modeling challenging. 

In our analysis, we have focused on one specific challenge: developing a statistical model with missing covariates in order to predict the level of platelet upon arrival at the hospital.
This model can aid creating an innovative response to the public health challenge of major trauma. The platelet is a cellular agent responsible for clot formation.  It is essential to control  its levels to prevent blood loss as quickly as possible in order to reduce early mortality in severely traumatized patients.  It is difficult to obtain the level of platelet in real time on arrival at hospital and, if available, its levels would determine how the patients are treated. Accurate prediction of this metric is thereby crucial for making important treatment decisions in real time.

We focus on patients after an accident who were sent directly to the hospital (not sent to Emergency Care Unit). 
After this pre-selection, 6384 patients remained in the data set. Based on clinical experience, in order to predict the level of platelet on arrival at the hospital, 15 influential quantitative measurements were included as pre-selected variables. Detailed descriptions of these measurements are shown in the supplementary materials \citep{sup}. These variables were included here because they were all available to the pre-hospital team, and therefore could be used in real situations.

Figure \ref{fig:per} shows the percentage of missingness per variable, varying from 0 to 60\%. If we were to perform the complete case analysis (\textit{i.e.}, ignoring all the observations with missing values)  only less than one third of the observations (1648 patients) would  still remain in the dataset. This loss of data demonstrates the importance of  appropriately handling the missing  values. 
\begin{figure}[!htbp]
\centering
\includegraphics[width=1\textwidth]{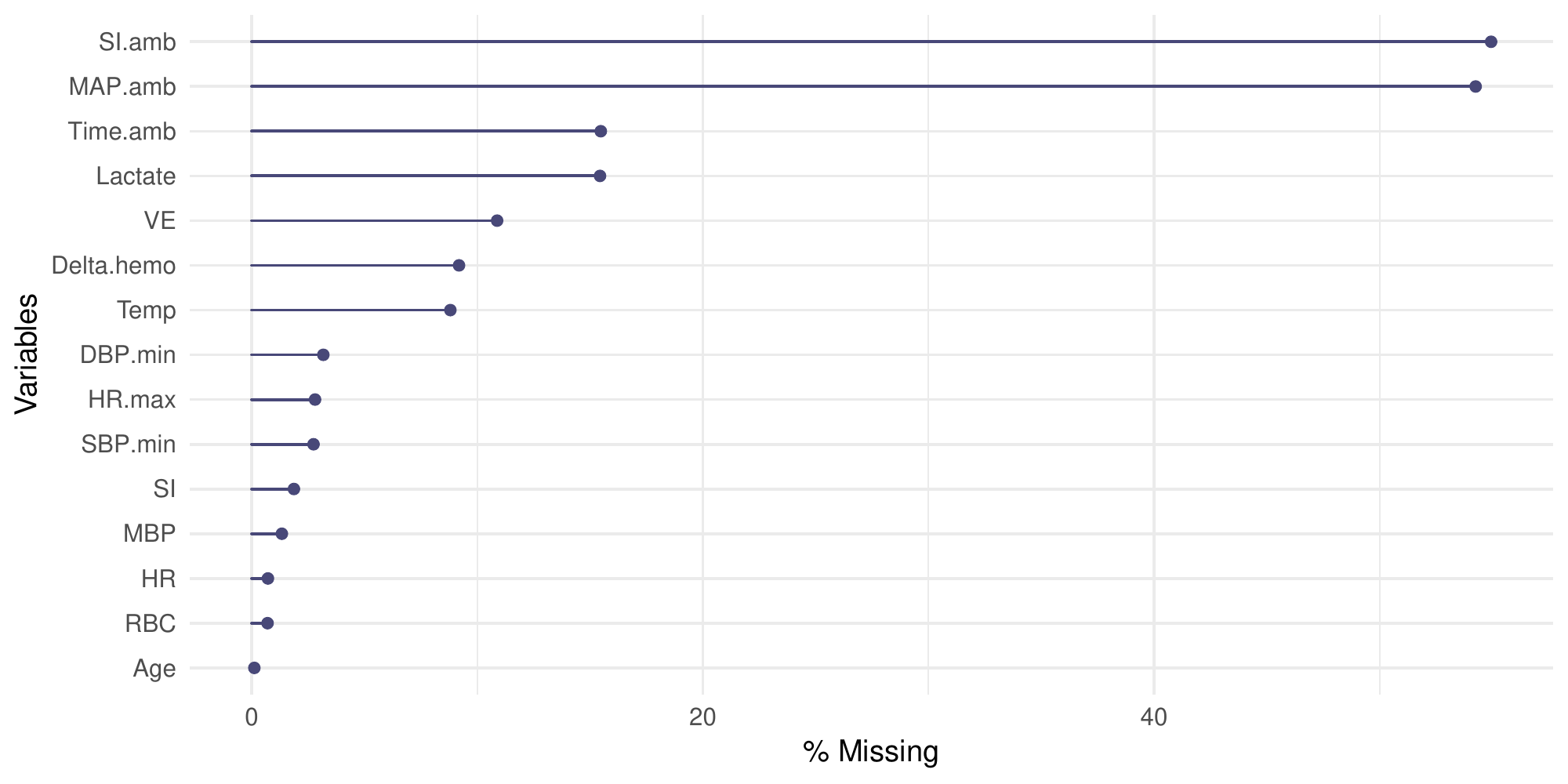}
\caption{Percentage of missing values in each pre-selected variable from TraumaBase.}
\label{fig:per}	
\end{figure}

\subsection{Model selection results}
As is customary in supervised learning, we divide the dataset into training and test sets. The training set contains a random selection of 80\% of observations whereas the test set contains the remaining 20\%. In the training set, we select a model and estimate the parameters.
We apply ABSLOPE and compare it with the same methods than those described in Section \ref{sec:simu}, namely MeanImp + SLOPE, MeanImp + LASSO, MeanImp + adaLASSO, MeanImp + SSL except ncLASSO since we do not known the sparsity and the $l_1$ bound of coefficients. Moreover, we also include: 
\begin{itemize}
\item BIC: Mean imputation followed by a stepwise method based on BIC;
\item RF: Mean imputation followed by a random forest \citep{rf}. This approach is assessed only for its prediction properties  as it does not explicitly select variables.
\end{itemize}
In the SLOPE type methods, we set the penalization coefficient $\lambda$ as BH sequence which controls the FDR at level $0.1$. Since we consider  our design matrix  being centered and without  an intercept, we also center the vector of responses and apply the procedure on  $\tilde{y} = y - \bar{y}$, where $\bar{y}$ is the mean of $y$.
We repeat the procedure of data splitting (into training and test sets) 10 times and Table \ref{tb:prediction} shows that, over 10 replications, how many times each variable is selected.  In addition,  Table \ref{tb:effet}  reports whether the selected variables by ABSLOPE have on average a positive or negative effect on the platelet. 

\begin{table}[t]
\footnotesize
\parbox{.68\linewidth}{
\vspace{1.3cm}
\caption{\label{tb:prediction} \small Number of times that each variable is selected over 10 replications. Bold numbers indicate which variables are included in the model selected by { ABSLOPE}.}
\vspace{0.4cm}
\centering
\fbox{
\begin{tabular}{lcccccr}%
  Variable  & ABSLOPE & SLOPE & LASSO & adaLASSO  & BIC\\ 
  \hline
 Age 					& \textbf{10} & 10 	&  4 & 10  & 10\\
 SI  					    & \textbf{10} 	& 2 	&  0 &   0  & 9\\
 MBP					& 1	             	& 10 	&  1 & 10 & 1\\
 Delta.hemo 		& \textbf{10} & 10 	&  8 & 10 & 10\\
 Time.amb   	 		& 2 	      		& 6 	&  0 &  4  & 0\\
 Lactate 				& \textbf{10} & 10 & 10 & 10 & 10 \\
Temp	 				& 2				& 10	&  0 &   0 & 0\\
 HR 						& \textbf{10}	& 10	&  1  & 10 & 10\\
 VE 						& \textbf{10} & 10	&  2  & 10 & 10\\
 RBC					& \textbf{10} & 10	& 10 & 10 & 10\\  
 SI.amb             	& 0   				&  0   &  0 & 0   & 0\\
 MBP.amb        	& 0   			&  0 &    0 & 0   & 0\\
 HR.max               & 3   			&  9 &    0 &  1 & 0\\
 SBP.min             & 5   				& 10 &  10 & 10 & 8\\
 DBP.min             & 2   			& 10 &   2 & 1  & 0
\end{tabular}
}
}
\hfill
\parbox{.28\linewidth}{
\caption{\label{tb:effet} \footnotesize The effect of the selected variables by {\it ABSLOPE} on the platelet.  ``$+$'' indicates positive effect while ``$-$'' negative; 0 indicates insignificant variables.}
\vspace{0.05cm}
\centering
\fbox{
\begin{tabular}{lccccr}%
  Variable  & Effect\\ 
  \hline
 Age 					& $-$ \\
 SI  					    & $-$ \\
 MBP 					& 0 \\
 Delta.Hemo 		& $+$ \\
 Time.amb 			& 0 \\
 Lactate 				& $-$ \\
 Temp 					& 0 \\
 HR 						& $+$ \\
 VE 						& $-$ \\
 RBC 					& $-$\\  
 SI.amb 				& 0 \\
 MBP.amb        	& 0 \\
 HR.max               & 0\\
 SBP.min             & 0 \\ 			
 DBP.min  			& 0
\end{tabular}
}
}
\end{table}

The TraumaBase medical team indicated that the signs of the coefficients were partially in agreement with their a-priori expectations. Large values of shock index, vascular filling, blood transfusion and lactate give signs of severe bleeding for patients and, thereby, lower levels of platelets. However, the effects of delta Hemocue and the heart rate on the platelet were not entirely in agreement with their opinion.

\subsection{Prediction performance}


In supervised learning, after  a model has been fitted on a training set, a natural step is to evaluate the prediction performance on a test set. Assuming  an observation $\bx=(\xobs,\xmis)$ in the test set, we want to predict the binary response $y$. One added difficulty is that the test set also contains missing values, since the training set and the test set have the same distribution (\textit{i.e.}, the distribution of covariates and the distribution of missingness). Therefore, we cannot directly apply  the fitted model to predict $y$ from an incomplete observation of the test $\bx$.

Our framework offers a natural remedy by marginalizing over the distribution of missing data,  given the observed ones.
More precisely, with $S$ Monte Carlo samples $(\xmis^{(s)}, 1 \leq s \leq S) \sim \dens(\xmis|\xobs),$
we estimate directly the response by maximum a posteriori value:
\begin{equation*}
\begin{split}
\hat{y} = \argmax_{y} \dens(y|\xobs) 
&= \argmax_{y} \int \dens(y|\bx) \dens(\xmis|\xobs) d \xmis\\
&= \argmax_{y} \mathbb{E}_{\dens_{\xmis|\xobs}} \dens(y|\bx)\\
&= \argmax_{y} \sum_{s=1}^S \dens \left(y|\xobs,\xmis^{(s)} \right).
\end{split}
\end{equation*}
Note that in the literature there are not many solutions to deal with the missing values in the test set \citep{consistency}. For those imputation based methods, we imputed the test set with mean imputation and predicted the platelet by $\hat{y} = X^{\rm imp} \hat{\beta}$. Finally we evaluate the relative $l_2$ prediction error:  $\text{err}=\frac{\Vert \hat{y} - y_{\rm } \rVert^2}{\Vert y_{\rm } \rVert^2}$. Prediction results obtained  are presented in Figure \ref{fig:boxplot}.
\begin{figure}[!htbp]
\centering
\includegraphics[width=0.65\textwidth]{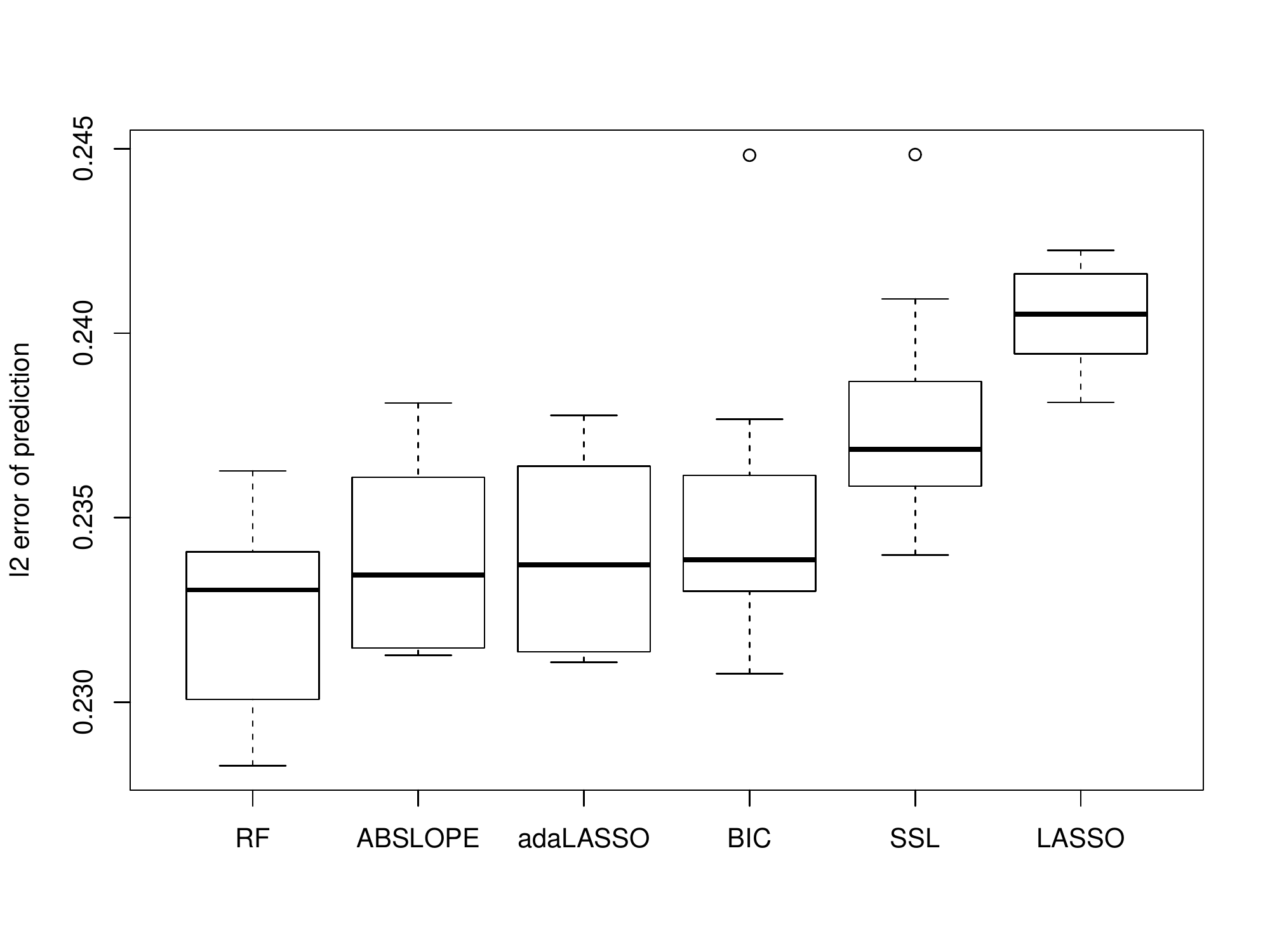}
\caption{Empirical distribution of prediction errors of different methods over 10 replications for the TraumaBase data. Results for SLOPE are not presented due to its large gap compared to others, with a mean of prediction error equals to 0.27.}
\label{fig:boxplot}	
\end{figure}

{ABSLOPE}'s performance is comparable to the one of Random Forest and adaptive LASSO, and is slightly better than   traditional stepwise regression and LASSO. There is a significant gap between the results of {ABSLOPE} and those of SLOPE. One of the possible reasons is that the classic version of SLOPE may encounter difficulties in the presence of correlation, while { ABSLOPE} works well  even with correlations (an aspect adopted from the Spike-and-Slab LASSO).  Random forests have excellent predictive capabilities which is consistent with the results of \citet{consistency} who show good performance of supervised machine learning even in the case of the simple mean imputation. However, it is difficult to interpret   results in terms of selected variables, which is often crucial for physicians. 

Figure \ref{fig:boxplot} and Table \ref{tb:prediction} show that  {ABSLOPE} and adaLASSO methods, which have the best predictive capabilities, select almost the same variables with adaLASSO selecting MBP (mean blood pressure) and {ABSLOPE} selecting SI (shock index). These two variables are highly correlated since both are measurements based on the systolic blood pressure.

Finally, we also performed ABSLOPE on the whole standardized dataset without cross-validation, and the formula of regression with model selection was reported as: $\rm{Platelet} = -6.92 \rm{ Age } - 7.28 \rm{ SI } + 6.53 \rm{ Delta.hemo } - 8.87 \rm{ Lactate } + 10.05 \rm{ HR }  - 3.96 \rm{ VE } - 8.91 \rm{ RBC } + 3.25 \rm{ SBP.min }$. This selection largely agrees with the results from cross-validation presented in Table \ref{tb:prediction}. The coefficient values demonstrate the importance of corresponding variables and provide a medical tool to predict the platelet value for a new patient.

\subsection{Results with Interactions}
We also consider a more complete model by adding second order interactions between the covariates, which increases the dimensionality at $p=55$. We apply the same procedure  as before and report the  predictive results  in Figure \ref{fig:boxplot_int}.

Table \ref{tb:pred_int} shows which variables are selected more than 5 times out of the 10 replications. Results for SSL and SLOPE are not presented due to their large gap compared to others, with a mean of prediction error equals to 0.35 and 0.40 respectively; BIC is not shown for this case with interactions, because it's computational heavy for this step-wise method with many variables. The average sizes of the variables set selected by ABSLOPE, LASSO and adaLASSO are respectively 6, 7 and 12.

\begin{table}[!htbp]
\small
\begin{minipage}[b]{0.5\linewidth}
\centering
\includegraphics[width=1\textwidth]{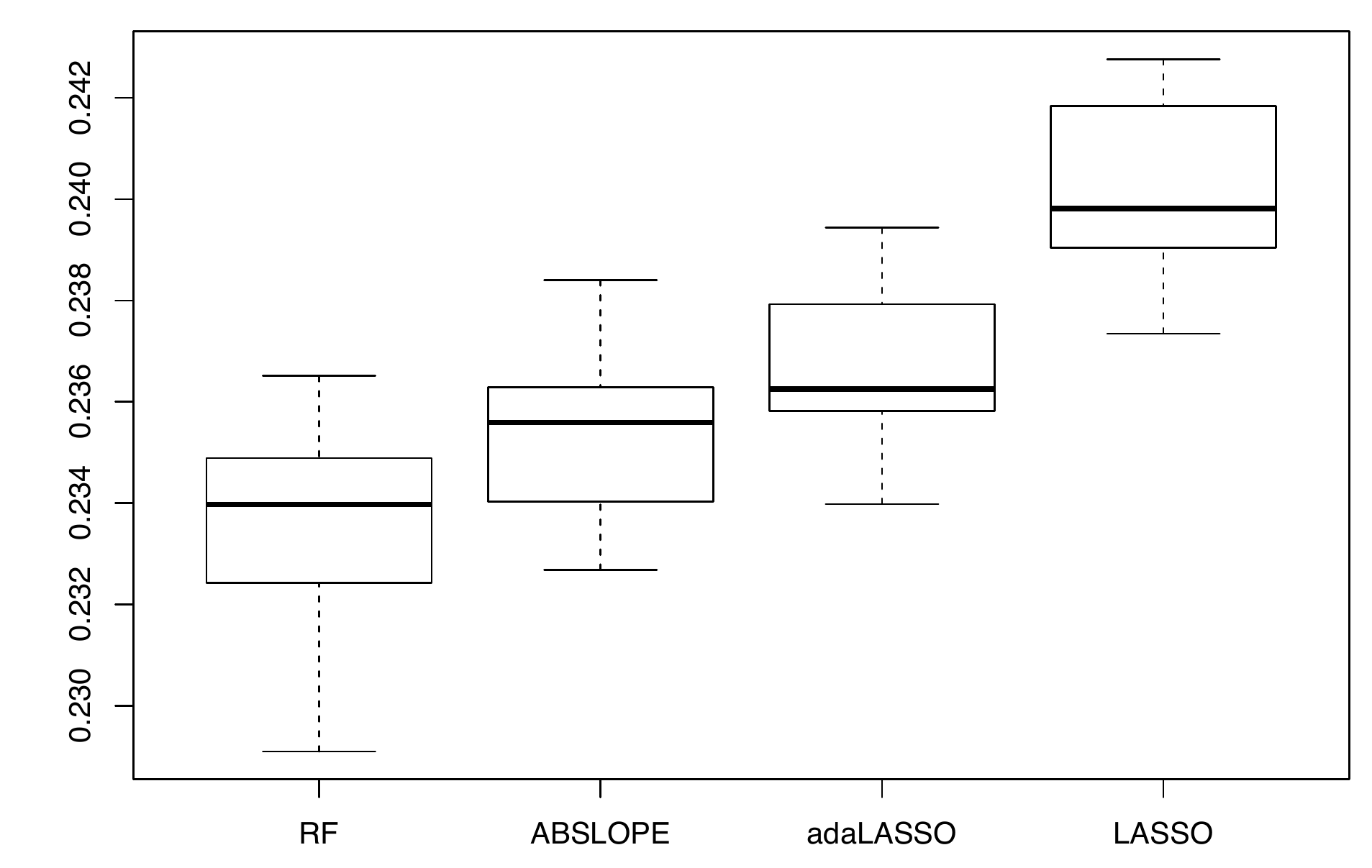}
\captionof{figure}{Empirical distribution of prediction errors of different methods over 10 replications for the TraumaBase data, with interactions between each pair of variables.}
\label{fig:boxplot_int}	
\end{minipage}\hfill
\begin{minipage}[b]{0.48\linewidth}
\renewcommand{\arraystretch}{0.7}
\centering
\scriptsize
\fbox{
\begin{tabular}{lccccr}%
  \textbf{Method} & \textbf{Variables selected} \\ 
  \hline
  ABSLOPE &  Age $*$ MBP.amb, Delta.hemo $*$ Lactate\\
  & Lactate $*$ RBC,  HR $*$ SBP.min \\
   \hline
   & RBC, \quad SBP.min, Age $*$ Lactate\\
  LASSO & Age $*$ VE, Delta.hemo $*$ Lactate\\
  & Delta.hemo $*$ VE , Lactate $*$ RBC \\
   \hline
   & Age $*$ Time.amb, Age $*$ HR\\
  & Age $*$ MBP.amb, Age $*$ SBP.min\\
  adaLASSO & MBP $*$ HR, Delta.hemo $*$ VE \\  
  & Lactate $*$ VE, HR $*$ HR.max\\
  & HR $*$ SBP.min, VE $*$ RBC
\end{tabular}
}
  \vspace{0.5cm}
\caption{\label{tb:pred_int} The variables selected more than 5 times out of the 10 replications, by each method.  ``$*$'' indicates the interaction between two variables.}
 \label{table:conf}
\end{minipage}
\end{table}

Again, {ABSLOPE} provides good results in terms of prediction while being sparse. We observe that when interactions are added, age often appears in combination with other variables. LASSO methods tend  to include a larger number of variables with a potentially increased false discovery rate. Note that the prediction properties with interactions are slightly worse than those without interactions, which happened due to the existence of missing values (\textit{e.g.} the interaction term between Age and Lactate will be missing if either of these two variables is unobserved).  In conclusion, other methods, apart from ABSLOPE, have a tendency to overfit when interactions are present.

\section{Discussion}\label{sec:concluded}
ABSLOPE penalizes noise coefficients more stringently to control for FDR while leaving larger effects relatively unbiased through an adaptive weighting matrix. In addition, casting our method within a Bayesian framework allows one to assign a probabilistic structure over models and estimate the pattern of sparsity. We develop an SAEM algorithm which  handles missing values and which treats model indicators as missing data. 
According to the simulation study,  ABSLOPE is competitive with other methods in terms of power, FDR and prediction error.
For future research,  we will consider the problem of high-dimensional model selection with missing values for categorical or mixed data and other missing mechanisms such as   MNAR.
\appendix

\section{Appendix}\label{app}

\subsection{Deviation of prior (\ref{eq:priorbeta}) started from SLOPE prior}\label{ann:dev}
We assume a random variable $z=(z_1,z_2,\cdots,z_p)$ has a SLOPE prior:
\begin{equation*}
\dens(z \mid \sigma^2;\lambda) \propto \prod_{j=1}^p  \exp\left\{-\frac{1}{\sigma}\lambda_{r(z,j)}|z_j| \right\},
\end{equation*}
and then define $\beta = W^{-1} z = (\frac{z_1}{w_1},\cdots,\frac{z_p}{w_p})$, or equally, $z_j = \beta_j w_j$  where the diagonal elements in the weight matrix are $w_j=c\gamma_j+(1-\gamma_j)=\begin{cases} c, & \gamma_j =1\\
1, & \gamma_j=0 \end{cases},  \quad j=1,2,\cdots,p.$
Then according to the transformation of variables, we have the prior distribution for $\beta$:
 \begin{equation*}
 \begin{split}
\dens(\beta\mid W,\sigma^2;\lambda) &\propto \left\lvert \det \left(\frac{dz}{d\beta}\right) \right\rvert \dens_z(W\beta \mid W, \sigma^2;\lambda) \\
&=\prod_{j=1}^p w_j \prod_{j=1}^p \exp\left\{-\frac{1}{\sigma} \lambda_{r(W\beta,j)}\lvert w_j \beta_j\rvert \right\}\\
&=
c^{\sum_{j=1}^p \iden(\gamma_j=1)}\prod_{j=1}^p \exp\left\{-w_j \lvert\beta_j\rvert\frac{1}{\sigma} \lambda_{r(W\beta,j)} \right\},
 \end{split}
\end{equation*}
which corresponds to our proposed prior (\ref{eq:priorbeta}).

\subsection{Missing mechanism}\label{ann:ignorable}
Missing completely at random (MCAR) means that there is no relationship between the missingness of the data and any values, observed or missing. 
In other words, for a single observation $X_i$, we have:
\begin{equation*} \label{eq:mcar}
\dens(r_i \mid y,\bx_i, \phi)=\dens(r_i \mid \phi)
\end{equation*}
Missing at Random (MAR), means that the probability to have missing values may depend on the observed data, but not on the missing data.
We must carefully define what this means in our case by decomposing the data $X_i$ into a subset $\xa_i$ of data that ``can be missing'', and a subset $\xb_i$ of data that ``cannot  be missing'', i.e. that are always observed.
Then, the observed data $\xiobs$ necessarily includes the data that can be observed $\xb_i$, while the data that can be missing $\xa_i$ includes the missing data $\ximis$. Thus, MAR assumption implies that, for all individual $i$,
\begin{equation} \label{eq:mar}
\begin{split}
\dens(r_i \mid y_i,\bx_i; \phi) &= \dens(r_i \mid  y_i,\xb_i; \phi) \\ &= \dens(r_i \mid y_i,\xiobs; \phi)
\end{split}
\end{equation}

MAR assumption implies that, the observed likelihood can be maximize and the  distribution of $r$ can be ignored \citep{little_rubin}. Assume that $\theta$ is the parameter to estimate. Indeed:

\begin{equation*} 
\begin{split}
\like(\theta,\phi ; y,\xobs,r)&= \dens(y,\xobs,r ; \theta,\phi) =\prod_{i=1}^n \dens(y_i,\xiobs,r_i ; \theta,\phi)\\
& = \prod_{i=1}^n \int \dens(y_i,\bx_i,r_i ; \theta,\phi)d\ximis\\
& = \prod_{i=1}^n \int \dens(y_i,\bx_i;\theta) \dens(r_i \mid \by_i,\bx_i;\phi)d\ximis,
\end{split}
\end{equation*}
then according to the assumption MAR (\ref{eq:mar}), we have:
\begin{equation*} 
\begin{split}
\like(\theta,\phi ; y,\xobs,r)&= \prod_{i=1}^n \int \dens(y_i,\bx_i;\theta) \dens(r_i \mid \by_i,\xiobs;\phi)d\ximis\\
& = \prod_{i=1}^n \dens(r_i \mid \by_i,\xiobs;\phi) \times \prod_{i=1}^n \int \dens(y_i,\bx_i;\theta)d\ximis\\
&= \dens(r\mid y,\xobs;\phi) \times \dens(y,\xobs; \theta) \\
\end{split}
\end{equation*}
Therefore, to estimate $\theta$, we aim at maximizing $\like(\theta; y,\xobs)=\dens(y,\xobs; \theta)$. So the missing mechanism can be ignored in the case of MAR.
\subsection{Standardization for MAR}\label{ann:std}
We update mean and standard deviation at each iteration of algorithm.
\begin{enumerate}
\item Initialization: In the initialization step, we first substitute missing values $\xmis$ with the mean of non-missing entries in each column, and obtain a imputed matrix $\tilde X^{0} = (\xobs, \xmis^0)$, where $\xmis^0$ contains imputed values. We denote the mean and standard deviation of each column of $X^{0}$, by the vectors $m^0$ and $s^0$ respectively. Then we centered and scaled the imputed $X^{0}$, s.t., for each observation $i$:
$$\hat{X}_i^{0} = (X_i^{0} -  m^0) \oslash (\sqrt{n} s^0),$$ where the $\oslash$ is used for Hadamard division. 
\item During $t^{\rm th}$ iteration of the algorithm, we obtain a new imputed dataset $X^{t}  = (\xobs, \xmis^t)$, where $\xmis^t$ contains imputed values in $t^{\rm th}$ iteration. 
Then we first reverse scaling using: $$\tilde{X}^{t}  = (\sqrt{n} s^{t-1}) \circ X^t + m^{t-1},$$ where the $\circ$ is used for Hadamard product. 
The vectors  $m^t$ and $s^t$ are then updated as the means and standard deviations of $\tilde{X}^{t}$. Finally we perform scaling on $\tilde{X}^{t}$ to obtain a scaled matrix:
$$\hat{X}_i^{t} = (\tilde{X}^{t}-  m^t) \oslash (\sqrt{n} s^t).$$
\end{enumerate}

\subsection{Details of the simulation step: sampling the latent variables}\label{ann:simu}
To perform the simulation step \eqref{eq:simusaem}, we use a Gibbs sampler. To simplify  notation, we hide the superscript, and note that all conditional distributions are computed  given the quantities from the previous iteration. 
\begin{enumerate}
\item Simulate $\gamma$. According to the dependency between variables presented in Figure \ref{fig:gm}, simulating the element $\gamma_j$ boils down to:
\begin{equation*}
\begin{split}
\gamma_j &\sim \dens(\gamma_j\mid\gamma_{-j},c,y,\xobs,\xmis,\beta,\sigma,\theta,\mu,\Sigma)\\
&=\dens(\gamma_j\mid\gamma_{-j},c,\beta,\sigma,\theta)\;,
\end{split}
\end{equation*}
where $\gamma_{-j} = (\gamma_1,\cdots,\gamma_{j-1},\gamma_{j+1},\cdots,\gamma_{p});$
\textit{i.e.}, sampling from a Binomial distribution with probability:
 \begin{equation}\label{eq:gamma}
\begin{split}&\mathbb{P}(\gamma_j=1\mid\gamma_{-j},c,\beta,\sigma,\theta) 
= \frac{\mathbb{P}(\gamma_j=1\mid\theta)\dens(\beta\mid\gamma_j=1,\gamma_{-j},c,\sigma)}{\sum_{\gamma_j \in \{0,1\}}\mathbb{P}(\gamma_j\mid\theta)\dens(\beta\mid\gamma_j,\gamma_{-j},c,\sigma)}\\
&= \left[ 1+\frac{(1-\theta)\exp\left(-{\frac{1}{\sigma}}\lvert\beta_j\rvert\lambda_{r(W^{0}\beta,j)}\right)  \times (c)^{\sum_{-j} \iden(\gamma_{-j}=1)}\prod_{-j}\exp\left( -w^{0}_{-j}\lvert\beta_{-j}\rvert\frac{1}{\sigma}\lambda_{r(W^{0}\beta,-j)} \right)}
{\theta c \exp\left(-c{\frac{1}{\sigma}}\lvert\beta_j\rvert\lambda_{r(W^{1}\beta,j)}\right)  \times (c)^{\sum_{-j} \iden(\gamma_{-j}=1)}\prod_{-j}\exp\left( -w^{1}_{-j}\lvert\beta_{-j}\rvert\frac{1}{\sigma^t}\lambda_{r(W^{1}\beta,-j)} \right)} \right]^{-1}\\
&= \left[ 1+\frac{(1-\theta)\exp\left(-{\frac{1}{\sigma}}\lvert\beta_j\rvert\lambda_{r(W^{0}\beta,j)}\right)   }
{\theta c \exp\left(-c{\frac{1}{\sigma}}\lvert\beta_j\rvert\lambda_{r(W^{1}\beta,j)}\right) } \times \frac{\prod_{-j}\exp\left( -w^{0}_{-j}\lvert\beta_{-j}\rvert\frac{1}{\sigma}\lambda_{r(W^{0}\beta,-j)} \right)}{ \prod_{-j}\exp\left( -w^{1}_{-j}\lvert\beta_{-j}\rvert\frac{1}{\sigma}\lambda_{r(W^{1}\beta,-j)} \right)}\right]^{-1},
\end{split}
\end{equation}
where the weighting matrix $W^{1}$ and $W^{0}$ have the same diagonal elements $w^{1}_{-j}=w^{0}_{-j} = 1-(1-c)\gamma_{-j}$, except for the position $j$: $w^{1}_{j} = c$ while $w^{0}_{j} = 1$.  Sampling from \eqref{eq:gamma} requires to store in memory ordered list which needs to be updated for every index $j$, such an approach could be computationally exhaustive. So we use an approximation,  which does not perturb solution significantly, by replacing both $W^{1}$ and $W^{0}$ by the estimate  of weighting matrix from previous iteration, noted by $W$. With the approximation, we partially retrieve the information of $\gamma_j$ from the last iteration, so the difference between the estimates from last and the current iteration will be reduced. Consequently,  (\ref{eq:gamma}) is drawn from:
 \begin{equation}\label{eq:gamma2}
\begin{split}
\mathbb{P}(\gamma_j=1\mid\gamma_{-j},c,\beta,\sigma,\theta, W) & =  \left[ 1+\frac{(1-\theta)\exp\left(-{\frac{1}{\sigma}}\lvert\beta_j\rvert\lambda_{r(W\beta,j)}\right)   }
{\theta c \exp\left(-c{\frac{1}{\sigma}}\lvert\beta_j\rvert\lambda_{r(W\beta,j)}\right) }\right]^{-1}\\
& =  \frac
{\theta c \exp\left(-c{\frac{1}{\sigma}}\lvert\beta_j\rvert\lambda_{r(W\beta,j)}\right)} {(1-\theta)\exp\left(-{\frac{1}{\sigma}}\lvert\beta_j\rvert\lambda_{r(W\beta,j)}\right)  + \theta c \exp\left(-c{\frac{1}{\sigma}}\lvert\beta_j\rvert\lambda_{r(W\beta,j)}\right) }\;,
\end{split}
\end{equation}
which can be interpreted as the posterior probability of binary signal indicator for $j^{\rm th}$ variable, given the prior guess $\mathbb{P}(\gamma_j=1\mid \theta) = \theta$ and the conditional likelihood of the vector $\beta$ given $\gamma_j=1$ and $\gamma_j=0$, see (\ref{eq:priorbeta}). 
\item Simulate $\theta$. The update of $\theta$ boils down to generate from:
\begin{equation*}
\begin{split}
\theta &\sim \dens(\theta\mid\gamma,c, y,\xobs,\xmis,\beta,\sigma,\mu,\Sigma, W)\\
 & =\dens(\theta\mid\gamma,\beta,\sigma,W) \propto \dens(\theta) \, \dens(\gamma\mid \theta)\;,
\end{split}
\end{equation*}
where $\dens(\gamma\mid \theta)$ is a Bernoulli distribution.
In addition, if we also assume a prior for $\theta$ as a Beta distribution $Beta(a,b)$ with $a$ and $b$ known, to offer additional initial information for the sparsity of signal, then the posterior is:
\begin{equation}\label{eq:theta}
Beta \left(a+\sum_{j=1}^p \iden (\gamma_j=1), b+\sum_{j=1}^p \iden(\gamma_j=0)\right)\;,
\end{equation}
from which we can generate the latent variable $\theta$.
The target distribution (\ref{eq:theta}) also takes the prior knowledge of the sparsity into consideration, for example:
\begin{itemize}
\item If $a = \frac{n}{100}$ and $b=\frac{n}{10}$, the prior mean on sparsity is 0.091, which has the same effect as a single observation;
\item If $a = \frac{2}{p}$ and $b = 1- \frac{2}{p}$, the prior mean on sparsity is $\frac{2}{p}$, which assumes a sparse structure a priori.
\end{itemize}
\item Simulate $c$. We also consider the weighting matrix $W$ from the previous iteration.
\begin{equation*}
\begin{split}
c &\sim \dens(c\mid\gamma,y,\xobs,\xmis,\beta,\sigma,\theta,\mu,\Sigma, W)\\
 & =\dens(c\mid\gamma,\beta,\sigma,W) \propto \dens(c) \, \dens(\beta\mid c, \gamma,\sigma, W)\\
&= p(c)\,c^{\sum_{j=1}^p\iden(\gamma_j=1)}\exp\left( -\frac{c}{\sigma}\sum_{j=1}^p \lvert \beta_j \rvert \lambda_{r\left(W\beta,j\right)}\iden(\gamma_j=1) \right)\;,
\end{split}
\end{equation*}
where $p(c)$ is the prior distribution of $c$. If the prior is chosen as $c \sim \mathcal{U}[0,1]$ then we just need to sample from a Gamma distribution truncated to [0,1]:
\begin{equation}\label{eq:c}
Gamma \left(1+\sum_{j=1}^p\iden(\gamma_j=1), \quad \frac{1}{\sigma}\sum_{j=1}^p \lvert\beta_j\rvert \lambda_{r(W\beta,j)}\iden(\gamma_j=1) \right)\;.
\end{equation}
If the signal is strong enough, \textit{i.e.}, $\beta_j$ is relative large compared to level of noise $\sigma$ when $\gamma_j = 1$, we will observe that the most typical values from the above Gamma distribution fall in the interval $[0, \, 1]$. As a result, the simulation will be closer to the original  Gamma distribution without truncation. However, if the signal strength go down, then the distribution will be more truncated and skewed towards 1, where $c$ exactly corresponds the inverse of average signal magnitude.
\end{enumerate}

\subsection{Proof of conditional distribution of missing data}\label{ann:xmislineq}
Proof of Proposition \ref{prop:xmislineq} is provided as follows.
\begin{proof}
For a single observation $x=(\xm,\xo)$  where $\xo$, and $\xm$ denotes observed and missing covariates respectively. 
Assume that $p(\xo,\xm; \Sigma, \mu)\sim \mathcal{N}(\mu,\Sigma)$ and let $y=x\beta+\varepsilon$ where $\varepsilon\sim N(0,\sigma^2)$. Then we have the following conditional distribution of the missing covariate with index $i$:
\begin{equation*}
\dens(\xm^i \mid \xo,y,\sigma,\beta, \Sigma,\mu,\xm^{-i})\propto \dens(\xo^i,\xm^i\mid \Sigma, \mu)\dens(y \mid \xo^i,\xm^i,\beta,\sigma)\;,
\end{equation*}
where $\xm^{-i} = \left( \xm^j, \, j \neq i \right).$ Denote $\mathcal{M}$ the set containing indexes for the missing covariates and $\mathcal{O}$ for the observed ones. We then explicitly give the formula, with $s_{ij}$ elements of $\Sigma^{-1}$:
\begin{equation*}
\begin{split}
& \dens(\xm^i \mid \xo,y,\sigma,\beta, \Sigma,\mu,\xm^{-i})  \propto 
\exp\left[-\frac{1}{2\sigma^2}(y-{x\beta})^2-\frac{1}{2} (x - \mu)^{\top} \Sigma^{-1} (x-\mu) \right]\\
\propto   &\exp \biggl[
-\frac{1}{2\sigma^2}\left(y-{\xo\beta_{\rm obs}} -  \xm^i \beta_i -\sum_{j \in \mathcal{M}, j\neq i}\xm^j \beta_j  \right)^2  \\
&\hphantom{\exp \biggl[} - \frac{1}{2}\left(
s_{ii}(\xm^i-\mu_i)^2+2\xm^i\sum_{j \in \mathcal{M}, j\neq i}(\xm^j-\mu_j)s_{ij}+2\xm^i\sum_{k \in \mathcal{O}}(\xo^k-\mu_k)s_{ik}\right) \biggl]\;.
\end{split}
\end{equation*}
After rearranging terms, with notations: $$m_i := \sum_{q=1}^p \mu_q s_{iq},\quad  u_i := \sum_{k \in \mathcal{O}}\xo^k s_{ik},\quad r := y-\xo{\beta_{\rm obs}}, \quad \tau_i :=\sqrt{s_{ii}+\frac{\beta_i^2}{\sigma^2}}\;, $$
we get:
\begin{equation}\label{eq:prop2eq1}
\begin{split}
 & \dens(\xm^i\mid \xo,y,\sigma,\beta, \Sigma,\mu,\xm^{-i})\\ \propto &
 \exp\biggl\{-\frac{1}{2}\biggl[{(\xm^i)}^2 \left(s_{ii}+\frac{\beta_i^2}{\sigma^2}\right)
 - 2 \xm^i \left( \frac{r\beta_i}{\sigma^2} +m_i-u_i \right) +2\xm^i\sum_{j \in \mathcal{M}, j\neq i} \left( \frac{\beta_i\beta_j}{\sigma^2}+s_{ij} \right)\xm^j\biggl]\biggl\} \\
\propto & \exp\biggl\{-\frac{1}{2}\biggl[{\xm^i}\tau_i 
 - \frac{r\beta_i /\sigma^2+m_i-u_i}{\tau_i} +\sum_{j \in \mathcal{M}, j\neq i}\frac{  \beta_i\beta_j/\sigma^2+s_{ij}}{\tau_i \tau_j}\xm^j \tau_j\biggl]^2\biggl\}\;.
 \end{split}
\end{equation}

By the other hand, we can conclude from equations (4.9) (4.10) in \citet{besag}, that, for $z=(z_i)_{i \in \mathcal{M}}$ where $z_i = \tau_i \xm^i$ we have:

\begin{equation}\label{eq:prop2eq2}
  \dens(z_i\mid \xo,y,\sigma,\beta, \Sigma,\mu,\xm^{-i})  \propto \exp \left[-\frac{1}{2}\left(z_i-\tilde{\mu}_{i}+\sum_{j \in \mathcal{M}, j\neq i} B_{i j}\left(z_{j}-\tilde{\mu}_{j}\right)\right)^{2}\right]\;,
\end{equation}
and
\begin{equation*}
 z \mid \xo, y; \Sigma,\mu,\beta,\sigma^2 \sim N(\tilde\mu, B^{-1})\;.
 \end{equation*}

Combine equations (\ref{eq:prop2eq1}) and (\ref{eq:prop2eq2}), we obtain the solution:
\begin{equation*}
 \frac{
r\beta_i /\sigma^2
+m_i-u_i}{\tau_i} 
-
  \sum_{j \in \mathcal{M}, j\neq i}\frac{
\beta_i\beta_j
  /\sigma^2+s_{ij}}{\tau_i\tau_j}\tilde\mu_j = \tilde \mu_i\;, \quad \text{for all }
i\in \mathcal{M}\;,
\end{equation*}
and 
\begin{equation*}
 B_{ij}= \begin{cases}  \frac{
 \beta_i\beta_j
 /\sigma^2+s_{ij}}{\tau_i\tau_j}, & \text{if  } i \neq j \\
  1, & \text{if  } i=j 
   \end{cases}, \quad \text{for all }
i, \, j \in \mathcal{M}\;.
\end{equation*}
\end{proof}

\subsection{Summary of algorithms}\label{ann:algo}
\begin{algorithm}[!htbp]
\caption{Solving ABSLOPE with SAEM.} 
\label{alg:ABSLOPE}
\begin{algorithmic}
\REQUIRE Initialization $\beta^0, \, \sigma^0, \, c^0, \theta^0, \, \xmis^{0}, \, \mu^0, \, \Sigma^0$;
\FOR{$t = 1,2,\cdots,\rm{Maxit}$} 
\STATE \textit{(Simulation step)}
\begin{enumerate}
\item  Generate $\gamma^t$ from (\ref{eq:gamma2});
\item  Generate $\theta^t$ from Beta distribution (\ref{eq:theta});
\item Generate $c^t$ from truncated Gamma distribution (\ref{eq:c});
\item  Generate $\xmis^t$ from Gaussian distribution (\ref{eq:xmislineq}); 
\end{enumerate}
\STATE \textit{(Stochastic Approximation step)}
\begin{enumerate}
\item Calculate $(\beta^t_{\rm MLE}$, $\sigma^t_{\rm MLE}$, $\mu^t_{\rm MLE}$,  $\Sigma^t_{\rm MLE})$, which are the MLE for complete-data likelihood integrating sampled missing values, as detailed in Subsection \ref{ssc:eta1};
\item With step-size $\eta_t = \begin{cases}1, & \text{if } t \le 20\\
\frac{1}{t-20}, & \text{if } t > 20
 \end{cases},$ update $$\beta^{t+1} \leftarrow \beta^{t} + \eta_t \left[ {\beta}^t_{MLE} -\beta^{t} \right]. $$
Update $\sigma$, $\mu$ and $\Sigma$ similarly;
\end{enumerate}
\IF{$ \lVert \beta^{t+1} - \beta^t \rVert ^2 < \rm{tol} $} 
\STATE \textbf{Stop};
\ENDIF 
\ENDFOR 
\ENSURE Estimates $\hat{\beta} \leftarrow \beta^t$ and indexes for model selection $\hat{\gamma} \leftarrow \gamma^t$
\end{algorithmic}
\end{algorithm}
We propose the ABSLOPE model and solve the problem of the maximization of the penalized likelihood using the SAEM algorithm, described in Algorithm \ref{alg:ABSLOPE}.
We also give the SLOBE algorithm in Algorithm \ref{alg:SLOB} which is an approximated and accelerated version.
\begin{algorithm}[!htbp]
\caption{SLOBE: a quick version of ABSLOPE.} 
\label{alg:SLOB}
\begin{algorithmic}
\REQUIRE Initialization $\beta^0, \, \sigma^0, \, c^0, \theta^0, \, \xmis^{0}, \, \mu^0, \, \Sigma^0$;
\FOR{$t = 1,2,\cdots,\rm{Maxit}$} 
\STATE \textit{(Imputation by expectation)}
\begin{enumerate}
\item \textbf{for} $j=1,2,\cdots,p$   \textbf{ do }  $\gamma^t_j$ $\leftarrow$ $\mathbb{E}(\gamma_j=1\mid\gamma_{-j},c,\beta,\sigma,\theta, W) $, according to (\ref{eq:slobgamma});
\item $\theta^t \leftarrow \mathbb{E}(\theta\mid\gamma,\beta,\sigma, W)$, according to (\ref{eq:slobtheta});
\item $c^t \leftarrow \mathbb{E}(c\mid\gamma,y,\xobs,\xmis,\beta,\sigma,\theta,\mu,\Sigma, W)$, according to (\ref{eq:slobc});
\item \textbf{for} $i=1,2,\cdots,n$   \textbf{ do }  $\ximis^t \leftarrow
\mathbb{E}(\ximis\mid y,\xiobs,\beta,\sigma,\mu,\Sigma),
$
according to Proposition \ref{prop:xmislineq};
\end{enumerate}
\STATE \textit{(Maximization of integrated likelihood)}
\begin{itemize}
\item $(\beta^{t+1}$, $\sigma^{t+1}$, $\mu^{t+1}$, $\Sigma^{t+1}) \leftarrow (\beta^t_{\rm MLE}$, $\sigma^t_{\rm MLE}$,  $\mu^t_{\rm MLE}$,  $\Sigma^t_{\rm MLE})$, which are the MLE for complete-data likelihood integrating the imputed missing values by expectation.
\end{itemize}
\IF{$ \lVert \beta^{t+1} - \beta^t \rVert ^2 < \rm{tol} $} 
\STATE \textbf{Stop};
\ENDIF 
\ENDFOR 
\ENSURE Estimates $\hat{\beta} \leftarrow \beta^t$ and indexes for model selection $\hat{\gamma} \leftarrow \gamma^t$
\end{algorithmic}
\end{algorithm}

\subsection{Initialization of ABSLOPE} \label{ann:init}
Here we suggest the following starting values:
\begin{itemize}
\item $\beta^0$ is obtained from elastic net LASSO \citep{glmnet}, or Spike and Slab LASSO \citep{spikeslablasso};
\item $\xmis^0$ are imputed by PCA (imputePCA)  \citep{missMDA}, or imputed by the mean of column (imputeMean);
\item $\mu^0$ and $\Sigma^0$ are estimated with the empirical estimators obtained from the imputed  initial data; 
\item $\sigma^0$ is given by the standard deviation: $\frac{\Vert y - \xmis^0 \beta^0 \Vert}{\sqrt{n - 1}}$;
\item $c^0= \min \left\{ \left(  \frac{\sum_{j=1}^p{\beta^0_j}}{\#\{j: \, \vert \beta^0_j \vert>0 \} +1}\right)^{-1} \sigma^0 \lambda_{r(\beta^0,1)}, 1 \right\}$, where the sign $\#$ means the cardinality of a set. $c^0$ can be interpreted as the inverse of average magnitude for the true signal, i.e, $\beta_j^0 \neq 0$; 
\item $\theta^0 = \frac{\#\{j: \, \vert \beta^0_j \vert>0 \}+a}{p+b}$ where $a$ and $b$ are known parameters of the prior Beta distribution on $\theta$. Here we choose \textit{i)} $a =\frac{2}{p}$ and $b =1 - \frac{2}{p}$, such that the prior mean on sparsity is $\frac{2}{p}$; \textit{ii)} $a =0.01n$ and $b =0.01n$; \textit{iii)} $a = 1$ and $b = p$. Our estimation results are not sensible to the choice of hyperparameters $a$ and $b$.
\end{itemize}

\section*{Supplementary Material}
\paragraph{package:} R-package ABSLOPE containing the implementation of the proposed methodology, available in  \citet{ABSl1}.
\paragraph{Codes:} Code to reproduce the experiments are provided in \citet{github}.
\paragraph{Additional supplementary materials:}  Some supplementary simulation results are presented in \citet{sup}.

\section*{Acknowledgment}
Wei Jiang was supported by grants from Région Île-de-France: \url{https:
//www.dim-mathinnov.fr}.  
The research of Malgorzata Bogdan was supported by the grant of the Polish National Center of Science Nr 2016/23/B/ST1/00454.
Veronika Rockova gratefully acknowledges support from the James S. Kemper Foundation Faculty Research Fund at the University of Chicago Booth School of Business.
We would like to thank Szymon Majewski for writing the code for { SLOBE}.
The authors are thankful for fruitful discussion with Edward I. George, Marc Lavielle,  Imke Mayer, Geneviève Robin and Aude Sportisse.

%
%
%
%

\bibliographystyle{apalike}

\bibliography{bibliography}
\end{document}